\documentclass[aps,prb,amsmath, amssymb, english, aps, prl, twocolumn,reprint,floatfix, superscriptaddress]{revtex4-1}

\usepackage[english]{babel}
\usepackage{graphicx}
\usepackage{verbatim}

\usepackage{epsfig}
\usepackage{epstopdf}
\usepackage{amsfonts}
\usepackage{amsthm}
\usepackage{amsmath}
\usepackage{amssymb}
\usepackage{color}
\usepackage[usenames,dvipsnames,svgnames,table]{xcolor}
\usepackage{hyperref} 
\hypersetup{backref,pdfpagemode=FullScreen,colorlinks=true,linkcolor=NavyBlue,citecolor=BrickRed}
\usepackage{makeidx}
\usepackage{pifont}

\usepackage{color}
\usepackage{grffile}


\def\expect#1{\mathinner{\langle{#1}\rangle}}

{\catcode`\|=\active
  \gdef\expect#1{\left<\mathcode`\|"8000\let|\bravert {#1}\right>}}
\def\bravert{\egroup\,\vrule\,\bgroup}

\def\beq{\begin{equation}}
\def\eeq{\end{equation}}
\def\be{\begin{equation}}
\def\ee{\end{equation}}

\def\epsk{\epsilon_{{\bf k}}}

\def\cG0{{\cal G}_0}

\def\spinup{\uparrow}
\def\spindown{\downarrow}

\def\a{\alpha}
\def\b{\beta}
\def\d{\delta}
\def\D{\Delta}

\def\eps{\epsilon}

\def\l{\lambda}

\def\s{\sigma}

\def\uc2{$U_{c2}$}
\def\uc1{$U_{c1}$}

\def\bea{\begin{eqnarray}}
\def\eea{\end{eqnarray}}
\def \bal{\begin{align}}
\def \eal{\end{align}} 
\def\#{\!\!}
\def\@{\!\!\!\!}

\def\+{\dagger}

\def\up{\spinup}
\def\down{\spindown}

\begin{document}

\title{\bf Enhancement of charge instabilities in Hund's metals by the breaking of rotational symmetry}
\author{Maria Chatzieleftheriou}
\affiliation{Laboratoire de Physique et Etude des Mat\'eriaux, UMR8213 CNRS/ESPCI/UPMC, Paris, France}

\author{Maja Berovi\'{c}}
\affiliation{International School for Advanced Studies (SISSA), Via Bonomea 265, I-34136 Trieste, Italy}

\author{Pablo Villar Arribi}
\affiliation{European Synchrotron Radiation Facility, 71 Av. des Martyrs, Grenoble, France}
\affiliation{Laboratoire de Physique et Etude des Mat\'eriaux, UMR8213 CNRS/ESPCI/UPMC, Paris, France}

\author{Massimo Capone}
\affiliation{International School for Advanced Studies (SISSA), Via Bonomea 265, I-34136 Trieste, Italy}
\affiliation{CNR-IOM-Democritos National Simulation Centre, UOS Trieste-SISSA, Via Bonomea 265, I-34136 Trieste, Italy}

\author{Luca de'~Medici}
\affiliation{Laboratoire de Physique et Etude des Mat\'eriaux, UMR8213 CNRS/ESPCI/UPMC, Paris, France}

\begin{abstract}
We analyze multi-orbital Hubbard models describing Hund's metals, focusing on the ubiquitous occurrence 
of a charge instability, signalled by a divergent/negative electronic compressibility, in a range of doping from the half-filled Mott insulator corresponding to the frontier between Hund's and normal metals. 
We show that the breaking of rotational invariance favors this instability: both spin-anisotropy in the interaction and crystal-field splitting among the orbitals make the instability zone extend to larger dopings, making it relevant for real materials like iron-based superconductors. 

These observations help us build a coherent picture of the occurrence and extent of this instability. 
We trace it back to the partial freezing of the local degrees of freedom in the Hund's metal, which reduces the allowed local configurations and thus the quasiparticle itinerancy. The abruptness of the unfreezing happening at the Hund's metal frontier can be directly connected to a rapid change in the electronic kinetic energy and thus to the enhancement and divergence of the compressibility.


\end{abstract}

\maketitle

\section{Introduction}

Materials with strong correlations between the valence electrons display very rich phase diagrams in which a variety of conventional and novel phases of matter compete and can be switched on and off via small changes of control parameters, as doping, temperature, pressure, strain.

A possible unifying principle behind the richness of correlated phase diagrams emphasizes the intrinsic weakness of the metallic state which makes it unstable in different channels. While the specific form of the broken-symmetry states can depend on material specifics  such as Fermi-surface nesting or other properties of the low-energy electronic states, identifying an intrinsic and general mechanism of instability which descends directly from strong electronic correlations would be a precious tool to explore the landscape of correlated materials.

These concepts have a long history which is strongly intertwined with high-temperature superconductivity. In particular, studies of the two-dimensional single-band Hubbard and $t-J$ models motivated by superconductivity in copper oxides have shown a tendency towards a divergence of the charge compressibility\cite{Emery_Kivelson-PhaseSep_tJ,Misawa_Imada-PhaseSep_Hubbard,Sorella_PhaseSeparation}, which leads to phase separation. Such instability can be the driving force behind the observations of charge-density waves and it has been proposed even as the trigger or a booster of the superconducting state\cite{CDG_PRL95}. In this frameworks phase separation has been indeed found in various decorations of the Hubbard model\cite{GrilliRaimondi_IntJModB,*CastellaniGrilliKotliar_2band_t-J,*Grilli_El-Ph,*Grilli_RCDK-PhaseSep_pdmodel,*Capone_PhaseSep_HubbardHolstein}.

More recently, a new page has been written after the discovery of iron-based superconductors (FeSC). These materials have indeed triggered the introduction of the new concept of  "Hund's metal". This name first appears in Ref. \onlinecite{Yin_kinetic_frustration_allFeSC} and highlights the fundamental role played by the intra-atomic exchange energy in shaping the metallic properties of these compounds and their degree of correlations. 

The properties of the Hund's metal and of the crossover which separates it from a more conventional metal have been discussed in a number of papers (Refs. contextually discussed in section \ref{sec:theory}). Among the distinctive features of these systems we find it worth to mention interaction-resilient metallic phases, orbital-selective correlations and anomalous magnetic properties.

Furthermore, Ref. \onlinecite{demedici_el_comp} reports the existence of a charge instability zone in the phase diagram of Hubbard-Hund models with different number of orbitals, which was later confirmed  in realistic DFT-based simulations of several iron-based superconductors (FeSC), like BaFe$_2$As$_2$, FeSe (both bulk and monolayer)\cite{VillarArribi_FeSe_el_comp} or their Chromium analogs\cite{Edelmann_Chromium_analogs}.
Remarkably, phase separation has been directly found by experiments in this family of compounds\cite{Civardi_Carretta-PhaseSeparation_Rb122}.

In this work we extend the picture of the charge instabilities of multi-orbital Hubbard models considering different interaction Hamiltonians and including perturbations such as crystal-field splitting and we find that Hamiltonians with a lower symmetry between orbitals display an enhanced tendency towards phase separation. More precisely the phase separation region is wider in doping for density-density interactions than for the rotational invariant  Kanamori and it increases as a function of the crystal-field splitting.

These new results help us identify the cause of the instability in multi-orbital models in terms of the quenching of the kinetic energy in the Mott insulating solution at half-filling, and in its sudden release at a doping, along the Hund's metal frontier. The extremal value of this frontier grows with the value of this quenched kinetic energy.

This paper is organized as follows. Sec.~\ref{sec:theory} introduces the main theoretical ideas, emphasizing the connection between the Hund's driven correlation and charge instabilities. In Sec. \ref{sec:models_methods} we introduce the models and the methodology. Secs.~\ref{sec:Kanamori_vs_Ising} and \ref{sec:cfs} present respectively results on the role of the symmetry of the interaction term and on the effect of a crystal-field splitting. Sec.~\ref{sec:interpret} discusses our interpretation of the results in terms of the kinetic energy of the system and in Sec.~\ref{sec:FeSe} we show that the larger instability zone found in the simulation for FeSe monolayer compared to the one for the bulk compound can be explained in terms of the enhanced crystal-field splitting found in the bi-dimensional case.
Conclusions and general remarks are in Sec.~\ref{sec:conclusions} while the Appendices~\ref{appendix:SSMF} and  \ref{appendix:KinEn_exp} report details on the slave-spin calculations illustrating the way the kinetic energy can be quenched and released depending on the degeneracy of the local many-body configurations, which is the mechanism at work highlighted in this paper. Appendix \ref{appendix:Hubbard_bands} shows how the width of the Hubbard bands is affected by this same mechanism.

\section{Hund's metals and charge instabilities}\label{sec:theory}

In this section we briefly review the main concepts defining a Hund's metal which are essential to build an understanding of the phase separation instability. 

When the description of a solid requires to use open-shell multi-orbital systems the theoretical modeling needs to include the atomic exchange coupling. The latter is often called Hund's coupling because it is responsible of the so-called Hund's rules, i.e., of the fact that the ground state configuration of a degenerate atom is the one where the total spin is maximal and, as a second condition, the orbital angular momentum is maximized.
This effect has now been taken into account in the treatment of magnetism and orbital order in insulating solids for many years, while the paramount influence on the conduction electrons in strongly correlated metals has been highlighted only recently\cite{Georges_annrev}.

In the standard band theory description the electronic many-body wave function is a simple (anti-symmetrized) product of individual Bloch functions, implying absence of correlations between the electron positions: in particular, no reduction of the probability for two or more electrons being close to one another is accounted for, besides the one implied by the Pauli principle. Expanding the wave function on a basis of local spin-orbitals this means that all possible local configurations (in which any number of electrons occupies any subset of the spin-orbitals at a given site) are realized, with a probability which is simply the product of the probability of each spin-orbital to be occupied in the single-particle Bloch functions.
For instance, for a set of degenerate spin-orbitals (this degeneracy being set by the point-group symmetry in the solid considered) at half-filling all possible configurations are realized with the same probability. Any splitting of this degeneracy, due to a reduction of the local crystal-field symmetry, will result in a different probability of occupation of the orbitals. This will indeed cause a different combined probability for the presence of electrons in different spin-orbitals, but always in an uncorrelated way, i.e., $\langle n_\a n_\b\rangle=\langle n_\a\rangle\langle n_\b\rangle$ (where $\a$ and $\b$ are any two local spin-orbitals, and $n_{\a}$, $n_{\b}$ are the corresponding number operators).


Interactions change this situation. Indeed the onsite Coulomb repulsion U penalizes  the local configurations with total occupancy far from the average density, compared to those closer to it, i.e., it reduces the onsite charge fluctuations. This blocking effect directly competes with the metallic behavior which is directly connected with free charge fluctuations. In the following we will discuss the outcome of this paradigmatic competition in the paramagnetic state,  assuming that no symmetry breaking takes place, i.e., that any ordering tendency is frustrated. 

Indeed metals are not immediately destroyed by small interactions as they can be described as Fermi-liquids even for fairly large interaction strength.
In a Fermi-liquid, the metallic character is maintained asymptotically at zero temperature for the excitations of lowermost energy, the quasiparticles. Their itinerancy is however reduced by the availability of configurations allowing the electrons to hop without an extra cost in energy. This depends both on the value of $U/t$ (where $t$ is the electron hopping amplitude in the system in absence of interactions) and on the filling of the system. When $U/t$ is large enough and the filling is commensurate (i.e., there is an integer number of electrons per site on average) the quasiparticles vanish and metallicity is lost, obtaining a Mott insulating state\cite{Brinkman_Rice,Georges_Krauth_92,rozenberg_mott_qmc}.
Irrespectively of the interaction strength, doping the system away from a commensurate filling necessarily induces extra sites with number of particles different from the average, which restores metallicity.

Turning to the role of multi-orbital effects, it has been shown that for models with $M$ local orbitals the critical interaction strength necessary for the Mott transition $U_c(M)$ increases linearly with the number of orbitals\cite{Lu_gutz_multiorb,Rozenberg_multiorb,Han_multiorb_Hund,Florens_multiorb}.

This is due to a subtle quantum effect: as we increase the number of orbitals, we have an increasing number of  local configurations with the same number of electrons, which remain degenerate if the model only features a Hubbard-$U$ repulsion which only measures the number of electrons per site.

On very general grounds, these configurations can combine in particular linear superpositions that have an increased hopping amplitude with respect to the bare atomic states. This implies that quasiparticles states containing these configurations have an enhanced kinetic energy allowing them to survive at larger $U$ compared to the single-orbital case. 

In the Appendices of this paper we illustrate explicitly these effects at half-filling (i.e., with a density of electrons per site $n=M$), a particular case in which analytic calculations are possible in the framework of one of the approaches used in this work, the Slave-Spin Mean-Field (SSMF).
Indeed in Appendix \ref{appendix:KinEn_exp} we show that thanks to this mechanism nearby the half-filled Mott insulator the kinetic energy \emph{per orbital} increases proportionally to a factor $M+1$. 

A similar argument holds also for the atomic-like charge excitations of energy ~$U$ from the ground state, which disperse more than in the one-band case , forming thus wider "Hubbard bands"\cite{gunnarsson_multiorb}. In Appendix \ref{appendix:Hubbard_bands} we show that in our description this effect is caused by the same factor increasing the kinetic energy.

This brings us to the crucial role of the Hund's exchange in this picture which turns out to be a reduction of this "extra" multi-orbital kinetic energy\cite{Koga_OSMT_2005}.
Hund's coupling $J$ splits the local states in energy, in particular favoring the high-spin over the low-spin ones and for each total spin the high-angular momentum ones.
This causes a reduction of the atomic degeneracy, reducing the allowed hopping processes and thus the gain in kinetic energy associated with coherent superpositions of the atomic states\cite{demedici_Vietri}.
In Appendix \ref{appendix:KinEn_exp} we estimate this reduction factor at half-filling as $E_{kin}(J=0)/E_{kin}(J)\gtrsim M+1$.
Again, in Appendix \ref{appendix:Hubbard_bands} we show that the same reduction applies to the width of the Hubbard bands in presence of a finite $J$.

These effects, together with the fact that the $J$ also contributes to the distance in energy with the configurations of filling different from average, tune the U$_c$ for the Mott transition (Appendix \ref{appendix:KinEn_exp}). In particular at half-filling, since J widens this gap in addition to U\cite{vanDerMarel_Sawatzky_MottHund} (so that one can define an effective Coulomb repulsion $U_{eff}=U+(M-1)J$), these effects collaborate to reduce the critical coupling compared to the J=0 case for every value of M, U$_c(M,J)<$ U$_c(M,J=0)$.\cite{Han_multiorb_Hund,pruschke_Hund,Ono_multiorb_linearizedDMFT,demedici_MottHund}

In this work we show that this quenching of the multi-orbital "extra" kinetic energy at half-filling causes the charge instability zone of the Hund's metal phase diagram reported in Ref. \onlinecite{demedici_el_comp} where the homogeneous metal is unstable towards phase separation or charge-ordered states. This zone - that could be of importance for high-$T_{c}$ superconductivity - was shown\cite{demedici_el_comp} to exist in Hubbard-Hund models with two, three and five orbitals and in realistic first-principle-based simulations of several iron-based superconductors (FeSC), like BaFe$_2$As$_2$, FeSe (both bulk and monolayer)\cite{VillarArribi_FeSe_el_comp} and also Chromium-based counterparts\cite{Edelmann_Chromium_analogs}.

These realistic descriptions feature all the important aspects of model calculations. 
In particular a non-zero Hund's coupling determines  two main zones in a phase diagram defined by $U$ and the filling: at small $U$ and filling far from half a moderately correlated metal; at large $U$ and filling closer to half a much more correlated "Hund's" metal\cite{Werner_SpinFreezing,Ishida_Mott_d5_nFL_Fe-SC,demedici_Janus}.
Upon crossing the frontier and entering the Hund's metal zone correlations increase and high-spin configurations dominate this paramagnetic metallic phase. Moreover for inequivalent orbitals the correlation strength becomes orbital selective \cite{Ishida_Mott_d5_nFL_Fe-SC,demedici_OSM_FeSC,Shorikov_LaFeAsO_OSMT,*Aichhorn_FeSe,Yin_kinetic_frustration_allFeSC,*Werner_122_dynU,*Misawa_d5-proximity_magnetic,*Backes_KRbCs122,*Bascones_FeSC_Magn_Review,*Si_NatureReview} and even orbital-selective Mott phases can happen, depending on the filling and hopping structure\cite{demedici_3bandOSMT,demedici_MottHund,demedici_OSM_FeSC}. 

In general terms, the frontier between the Hund's and the standard metal is the place where a charge instability is found.
However its extension in the $U$-doping plane turns out to be strongly dependent on the specific system at hand.  
Orbitally symmetric models with featureless semi-circular densities of states (DOS) were investigated first\cite{demedici_el_comp}. There, it was found that the shape and extension of the instability zone - marked by diverging and negative electronic compressibility - is different depending on the number of orbitals in the model. Indeed, for increasing $M$ it spans smaller $U$-ranges but larger doping ranges. 
Moreover,  in realistic simulations of 5-orbital materials the instability zone can extend even as far as 1 electron (or 1 hole) of doping away from half-filling, i.e., into the region relevant for the stoichiometric Fe-based superconductors, and actually correlates positively with the experimental superconducting $T_{c}$ in the cases investigated thus far\cite{demedici_el_comp,Edelmann_Chromium_analogs,VillarArribi_FeSe_el_comp} . 

As a matter of fact, however, the instability zone still varies from compound to compound, and one of the goals of this paper is to clarify some of the trends found in material simulations through a model study, at the same time identifying the physical mechanism behind these trends.

In particular, we here show that the diverging/negative compressibility zone in the phase diagram:
\begin{itemize}
\item is wider in doping for density-density interactions than for the rotational invariant standard Kanamori form. In both cases however we find that the maximum doping reached grows with the number of orbitals $M$,
\item is wider in doping when the orbitals are not degenerate,  the larger the crystal-field  splitting, the larger the doping range.
\end{itemize}

\section{Model and Methods}\label{sec:models_methods}

We analyze a general multi-band Hubbard model with $M=2,3$ and $5$ orbitals, of which the Hamiltonian reads $\hat H-\mu \hat N=\hat H_0+\hat H_{int}-\mu \hat N$ with:
\be
\hat H_0 = \sum_{i\neq jmm'\s} t^{m m'}_{ij} d^\dag_{im\s}d_{jm'\s}+\sum_{im\s}\eps_mn_{im\s},
\label{eq:H0}
\ee
where $d^\dag_{im\s}$ creates an electron with spin $\s$ in orbital $m=1,\ldots, M$ on site $i$ of the lattice, and  $n_{im\s}=d^\dag_{im\s}d_{im\s}$ is the number operator, $\hat N=\sum_{im\s}n_{im\s}$, is the total number of electrons. Any band structure can be written this way, but here we only consider the particular case of diagonal hopping in orbital space, equal for all orbitals i.e.,  $t^{m m'}_{ij}=t_{ij}\d_{mm^\prime}$. The chemical potential  $\mu$ sets the average density of electrons per lattice site $n=\langle \hat N \rangle/{\cal N}_{sites}$.

The interacting part of the Hamiltonian reads:
\begin{eqnarray}\label{eq:ham_kanamori} 
\hat H_{int}\,&&=U\sum_{im} \tilde n_{im\up} \tilde n_{im\down}\, +\,U^\prime\!\!\sum_{im\neq m'} \tilde n_{im\up} \tilde n_{im'\down}\,\\
&&+(U^\prime-J) \!\!\!\! \sum_{im< m',\sigma} \!\! \tilde n_{im\s} \tilde n_{im'\s} + \nonumber \\
&&-\a J\!\!\sum_{im\neq m'} \!\! d^+_{im\spinup}d_{im\spindown}\,d^+_{im'\spindown}d_{im'\spinup} \nonumber \\
&&+\a J\!\!\! \sum_{im\neq m'} \!\! d^+_{im\spinup}d^+_{im\spindown}\,d_{im'\spindown}d_{im'\spinup}, \nonumber
\end{eqnarray} where $\tilde n_{im\s}=n_{im\s}-1/2$ is a particle-hole symmetric form of the density operators and customarily\cite{Georges_annrev} we set $U^\prime=U-2J$. 
We will consider two cases for the interaction: the Kanamori form $H_{int}(\a=1)$ and its simplified density-density term-only version $H_{int}(\a=0)$.

We treat the $\a=1$ model using the Rotationally-Invariant Slave-Boson mean-field (RISB) which can correctly treat the off-diagonal interaction terms and the $\a=0$ model both with RISB and the Slave-Spin Mean-Field approximation (SSMF). The technical details of these methods are given respectively in Refs.~\onlinecite{Lechermann_RISB} and \onlinecite{demedici_Vietri}. The two methods are known to coincide exactly in specific cases (e.g., at particle-hole symmetry) but show some very small differences in general. In all we will expose here these small differences are irrelevant, yet we will underline which method was used for generating the data shown in each figure. 
We will explore the typical physical range of Hund's coupling $0<J/U<1/3$. When focusing on specific values we have chosen $J/U=0.12\div0.15$ for the $\a=1$ case, which is typically found in correlated materials with $3d$ and $4d$ transition-metals, and $J/U=0.25$, a customary value which for the $\a=0$ also reproduces well the physics of several materials (like FeSC) when described with the $\a=0$ model within SSMF\cite{demedici_Vietri}.

We focus on the normal, non-magnetic, metallic phase at zero temperature.

Both methods treat the model in a framework in which the aforementioned local configurations are handled by the auxiliary "slave" variables (respectively bosons or spin-$1/2$). The effect of the interaction on the local configuration and on their relative weight is embodied in a renormalization of the original hopping amplitudes for the low-energy states of the system, so to describe Fermi-liquid quasiparticle excitations through the effective Hamiltonian:
\be\label{eq:QP_ham}
H_f-\mu N= \sum_{km\s} (Z_m\eps_k+\eps_m-\lambda_m-\mu) f^\+_{km\s}f_{km\s}
\ee
where $f^\+_{km\s}$ is the creation operator of a quasiparticle with momentum $k$, orbital (band) character $m$ and spin $\s$ and $\epsk$ is the bare electronic dispersion relation which is the same for all the bands. We can characterize it by its density of states, and we customarily choose a semi-circular DOS $D(\eps)\equiv\frac{2}{\pi D}\sqrt{1-(\frac{\eps}{D})^2}$ of bandwidth $W=2D$ for all bands. 
For this particle-hole symmetric DOS and for the particle-hole symmetric form of $H_{int}$ obtained when expressed in term of the $\tilde n_{im\s}$, $\mu=0$ guarantees half-filling of the bands (in absence of crystal-field splitting, or when the splitting is itself particle-hole symmetric).

In the following section we  study the evolution of the zones of enhanced/divergent compressibility of the electronic fluid 
\begin{equation}
\kappa_{el}=\frac{dn}{d\mu}
\end{equation}
induced by Hund's coupling, as a function of the type of interaction, the crystal-field splitting and the number of orbitals in the model.

\section{Kanamori vs density-density interaction}\label{sec:Kanamori_vs_Ising}
	
\begin{figure*}[h!]
\begin{center}
\includegraphics[width=\textwidth]{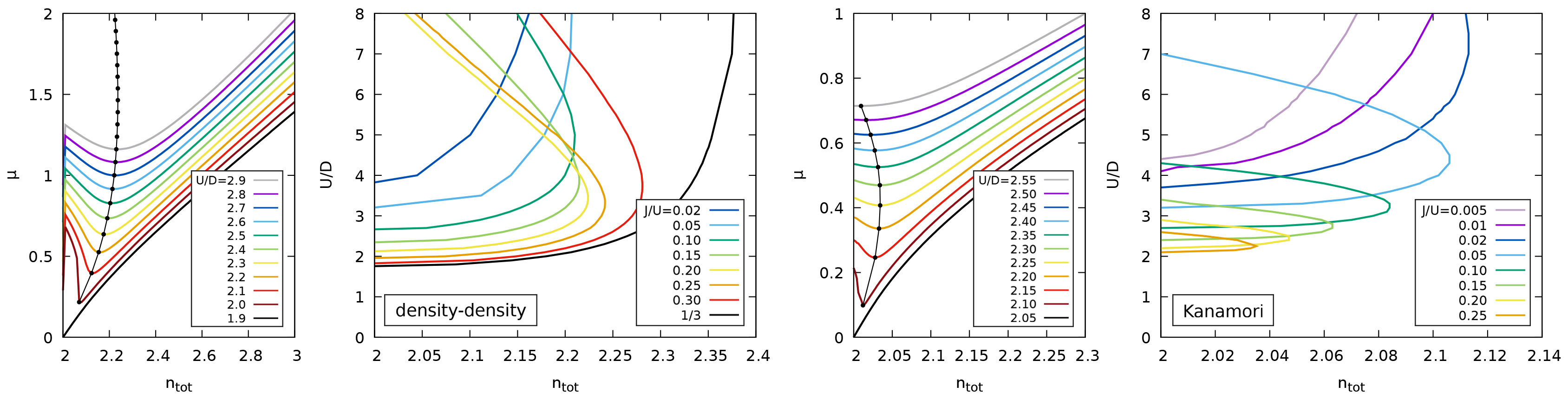}
\caption{2-orbital Hubbard model. The two leftmost panels show, for the model with density-density interaction ($\a=0$ in eq.~\ref{eq:ham_kanamori}) density vs chemical potential curves (left) for a typical value of the Hund's coupling relative strength $J/U=0.25$ and the evolution of the instability zone in the interaction-density plane, for different values of $J/U$. The unstable zone is delimited by the colored curves and the $y$-axis. Calculations performed within the Slave-Spin Mean-Field scheme (SSMF). Rightmost panels: same for the model with Kanamori interaction ($\a=1$), calculated within the Rotationally-Invariant Slave-Bosons mean-field scheme (RISB). Adapted from Ref.~\onlinecite{demedici_el_comp}.}
\label{fig:2b_Kanamori_Ising}
\end{center}
\end{figure*}	

In this section we compare the different extension of the instability zone between the models with Kanamori ($\a=1$) and density-density ($\a=0$) interactions. This can help comparing two different interaction Hamiltonians which are both used in studies of models and materials and to highlight the role of rotational invariance in the interaction.

We start  with the results for a 2-orbital model, which have already been shown  in the supplementary material of Ref.~\onlinecite{demedici_el_comp} and here are featured in Fig.~\ref{fig:2b_Kanamori_Ising}. Here we report the boundary of the zone of negative compressibility in the plane of density and interaction for a number of different values of $J/U$. The compressibility, as visible from the typical $\mu(n)$ curves reported in the figure, is found to be positive and well behaved outside this zone; it diverges on the frontier, and is negative between the frontier and the $n=2$ axis. The lowest border of the frontier departs from the $U_{c}$ where a Mott transition is realized at half-filling.

Even if the evolution with $J/U$ shows differences among the two cases, a clear trend is obviously visible: at each value of $J/U$ the model with density-density interaction has a more extended instability zone. In particular the instability extends in a larger range of densities.

\begin{figure*}[h!]
\includegraphics[width=\textwidth]{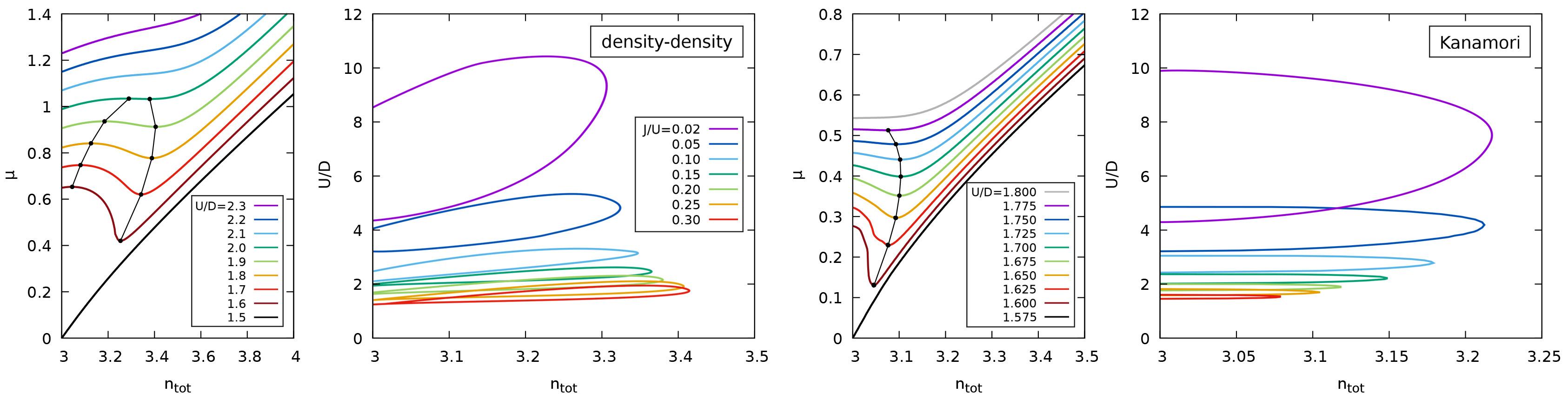}
\caption{Same as in Fig.~\ref{fig:2b_Kanamori_Ising}, but for the 3-orbital Hubbard model. Calculations performed within RISB mean-field. The color code for $J/U$ is the same in both phase diagrams.}
\label{fig:3b_Kanamori_Ising}
\end{figure*}

\begin{figure*}[h!]
\begin{center}
\includegraphics[width=\textwidth]{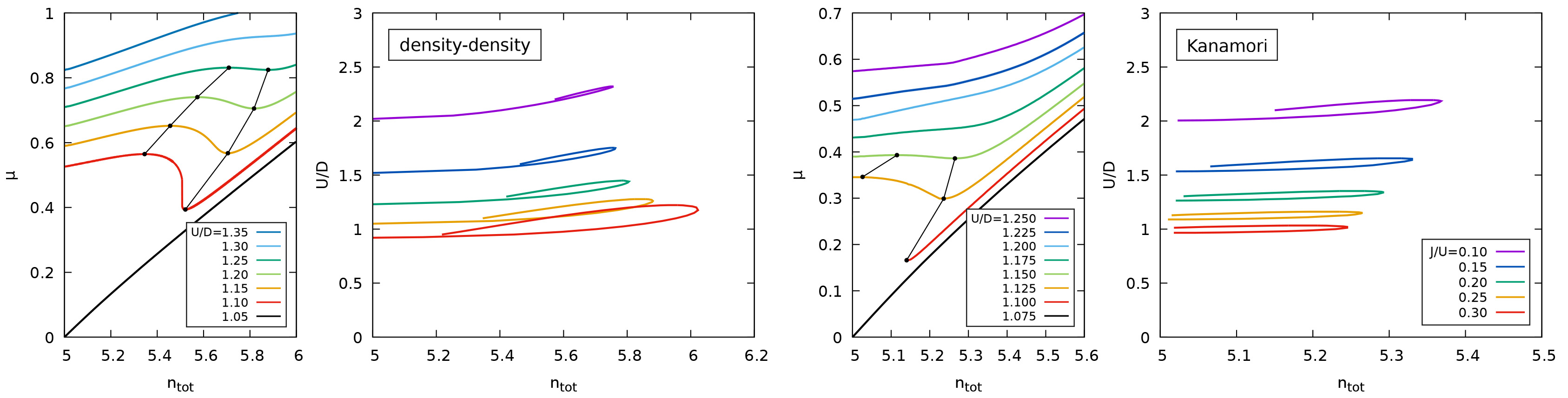}
\caption{Same as in Fig.~\ref{fig:2b_Kanamori_Ising}, but for the 5-orbital Hubbard model. Calculations performed within RISB mean-field.The color code for $J/U$ is the same in both phase diagrams.}
\label{fig:5b_Kanamori_Ising}
\end{center}
\end{figure*}

This trend is confirmed both in the 3-orbital and in the 5-orbital models studied in this work, for which we study the $\a=1$ case for the first time, and compare it to the $\a=0$ case. In Fig.~\ref{fig:3b_Kanamori_Ising} and ~\ref{fig:5b_Kanamori_Ising} the curves $\mu(n)$ are reported for selected values of $U$ and $J/U$, and the density-density model always shows a noticeably more extended instability zone.

Interestingly the range of doping from half-filling, $\d=n-M$, for which the instability is found increases with the number of orbitals $M$.

We can conclude that, as in the case of density-density interaction ($\a=0$) studied in Ref. \onlinecite{demedici_el_comp} the zone of instability spans a larger and larger doping range the larger is the number M of orbitals in the model. We also confirm that in all cases the $\a=0$ model has a larger instability zone than the corresponding $\a=1$ model, thus the breaking of the rotationally invariance of the interaction enhances the instability region as a function of doping.

\section{Extension of the instability zone with crystal-field splitting}\label{sec:cfs}

We now focus on the dependence of the phase separation instability on the crystal-field splitting between different orbitals. This term obviously reduces the symmetry of the model by breaking the orbital degeneracy at the single-particle level.

We start from the 2-band model, where the only possible splitting is given by the energy difference between the two orbitals  $\D=\eps_1-\eps_2$. In our model where all bands have identical bandwidth and DOS a finite $\D>0$ implies a difference in band/orbital populations $n_2>n_1$.
However, at half-filling this does not break particle-hole symmetry, i.e., holes in one of the bands behave like electrons in the other one and the two bands still show identical physics. In particular, they undergo a common Mott transition as a function of $U$. 
\begin{figure}[h!]
\begin{center}
\includegraphics[width=0.5\textwidth]{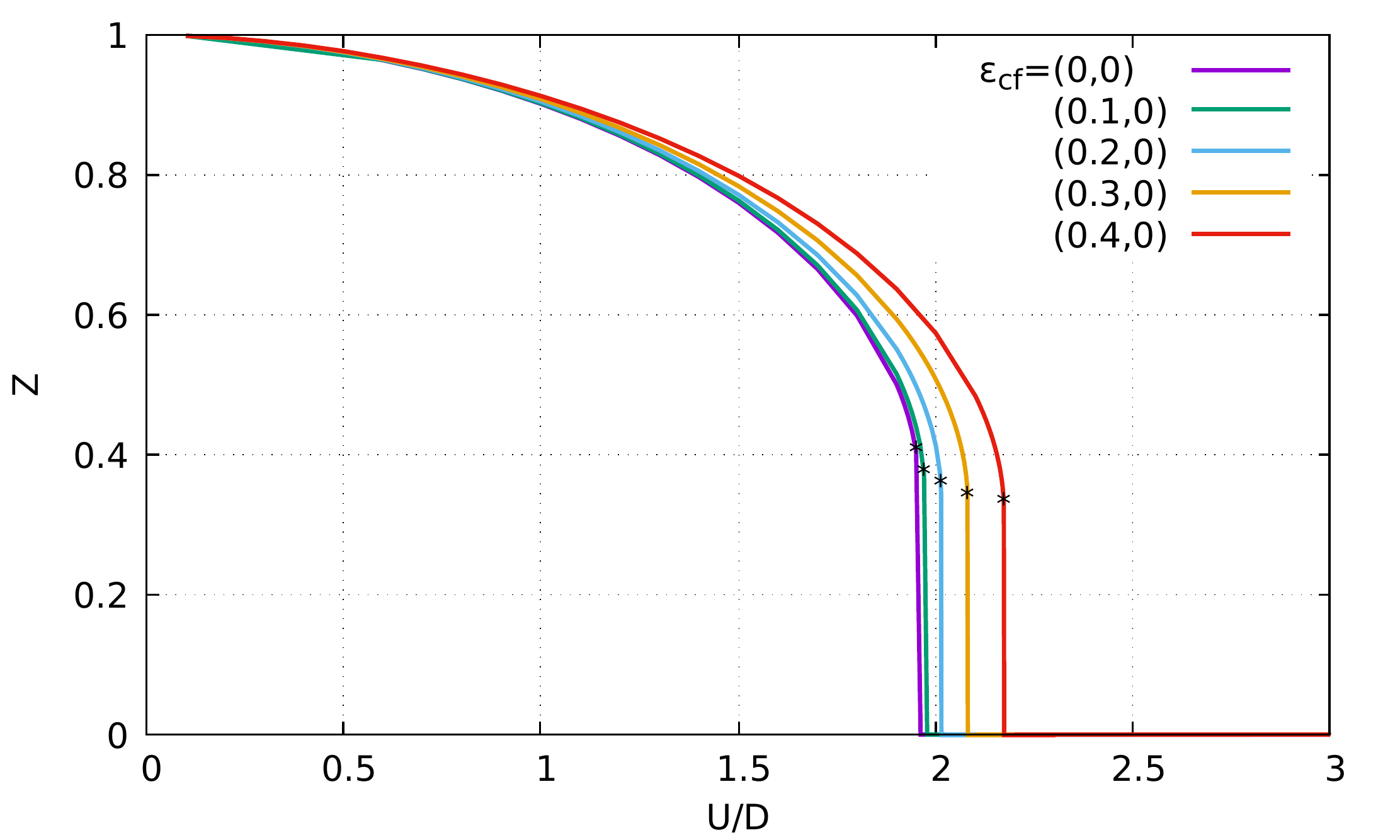}
\caption{Quasiparticle renormalization factor $Z$ (inverse mass enhancement) as a function $U/D$ in the 2-orbital Hubbard model with density-density interaction and Hund's coupling relative strength $J/U=0.25$. The stars mark the value of $Z$ for $U\rightarrow U_c$. Calculated within SSMF.}
\label{fig:2b_cfs_Z_vs_U}
\end{center}
\end{figure}

The net effect of the crystal-field splitting on the Mott transition is to raise the critical interaction strength U$_c$ needed to get the Mott insulator.
This is easily understood in terms  of the Mott gap $E_G = E(3) + E(1) -2E(2)$, where $E(N)$ is the energy of the atomic ground state with N particles. 
Indeed in the 2-orbital model the energy of the local high-spin configurations with 2 particles $E(2)$   is untouched by a symmetric crystal-field splitting, while half of the configurations with 3 or 1 particles are lowered in energy, thus diminishing the Mott gap with respect to the case with $\D=0$.
In the slave-spin formalism this is easily shown\cite{demedici_Vietri} to tune also the low-energy renormalization and one can analytically solve for the $U_c=4\bar\eps_0/(1+j)(1+\sqrt{1+(\D/4\eps_0)^2})$, for a chosen value of the fixed ratio $j=J/U$. This is an approximation to the trend seen in Fig.~\ref{fig:2b_cfs_Z_vs_U} because it is calculated in perturbation theory for a vanishing $Z$, which holds if a second order Mott transition is realized. Here instead, as visible in the figure, $Z$ has a jump at the transition meaning that it is actually of the first order. Nevertheless, the analytic result can be taken as a guidance for the trend of the 1st order transition. This result is also confirmed by computationally heavier and more accurate  Dynamical Mean-Field Theory (DMFT) \cite{Werner_high-low_Hund}.

\begin{figure}[h!]
\begin{center}
\includegraphics[width=0.47\textwidth]{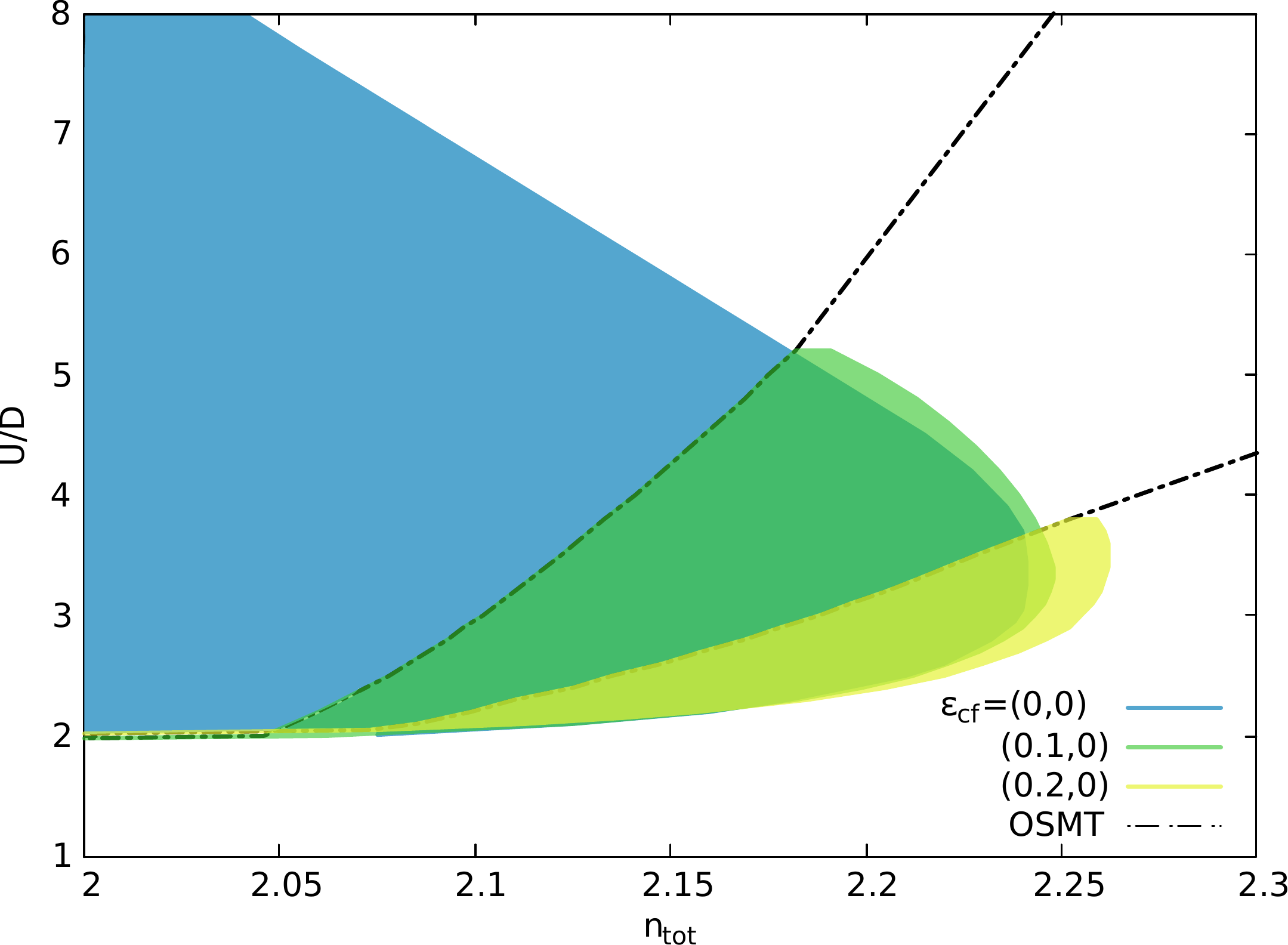}
\caption{ Phase diagram of the 2-band Hubbard model with density-density interaction ($\a=0$, $J/U=0.25$) and crystal-field splitting $\D/D=0, 0.1, 0.2$ calculated within Slave-Spin Mean-Field (SSMF). For doping below each dash-dotted line the system is in an orbitally-selective Mott phase (OSMP) with the band higher in energy insulating and the other metallic. The zones limited by the colored lines and the dash-dotted lines are unstable towards phase separation (negative compressibility, diverging at the border with the correlated multi-band metal at large doping).}
\label{fig:2b_cfs_SSMF}
\end{center}
\end{figure}

\begin{figure}[h!]
\begin{center}
\includegraphics[width=0.47\textwidth]{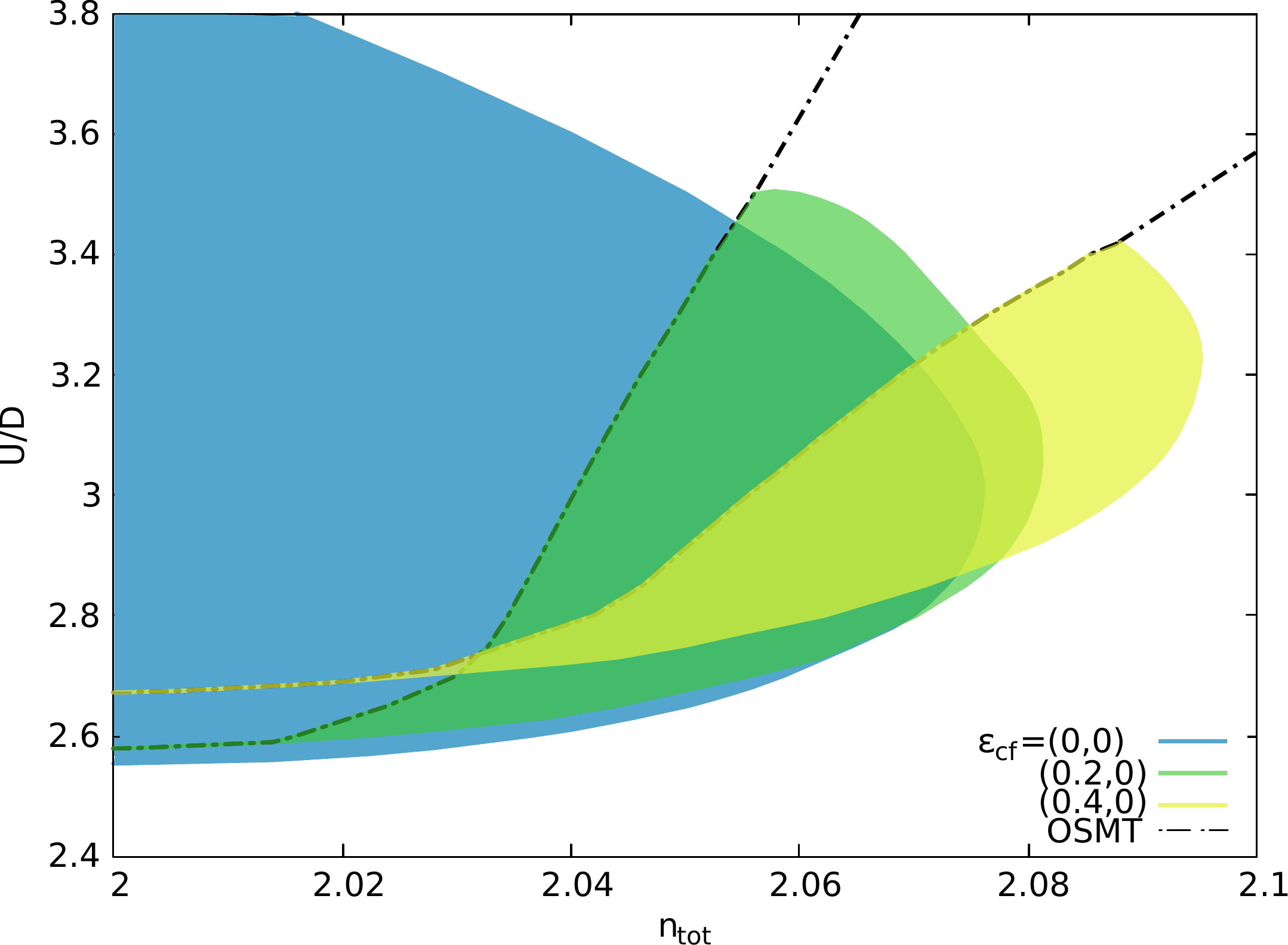}
\caption{Same as Fig.~\ref{fig:2b_cfs_SSMF} (2-band Hubbard model) but with Kanamori interaction ($\a=1$, $J/U=0.12$) and crystal-field splitting $\D/D=0, 0.2, 0.4$, calculated within Rotationally-Invariant Slave-Bosons mean-field (RISB).} 
\label{fig:2b_cfs_RISB}
\end{center}
\end{figure}

For $U>U_{c}$, for every finite doping the breaking of the orbital symmetry due to a finite $\D$ introduces a difference in the orbital behaviour. In particular the population is different among the orbitals, and the degree of electronic correlation associated to each orbital follows this difference, due to the emergent "orbital decoupling"\cite{demedici_3bandOSMT,demedici_MottHund,demedici_OSM_FeSC,demedici-SpringerBook}. This mechanism triggered by Hund's coupling in proximity to a half-filled Mott insulator decouples the charge excitations in the different orbitals, rendering the Mott physics and the degree of correlation of the orbitals almost independent of each other.
In this situation we can find orbital-selective Mott transitions (OSMT) where some orbitals become Mott localized while others remain metallic.
In our model indeed, for any finite $\D$ positive doping populates first the band lower in energy, which then becomes metallic, while the other remains half-filled and Mott insulating\cite{Werner_high-low_Hund}. This is the orbital-selective Mott phase (OSMP) of which the frontier is marked in Fig.~\ref{fig:2b_cfs_SSMF} and ~\ref{fig:2b_cfs_RISB} by dot-dashed lines.  

In the two figures we see that the orbital-selective transition line seems  to "chop" the zone of negative compressibility which in this model occupies a large zone for $U>U_{c}$ and a range of doping around half-filling. However if the doping is further increased, both orbitals turn metallic and the OSMP turns into a two-band metal which is still unstable towards phase separation in the vicinity of the transition. In this sense, we can conclude that the  OSMT  "chops and pushes" the instability zone, so that for large $\D$ it takes rather the shape of a slice of the original zone, moved towards somewhat higher doping.

The same effect can be seen both in the $\a=0$ (Fig.~\ref{fig:2b_cfs_SSMF} and in the $\a=1$ (Fig.~\ref{fig:2b_cfs_RISB}) cases. 
The main difference is that in the former the crystal field seems more effective in chopping than pushing the instability towards higher dopings, and the inverse happens in the latter instead.
For all values of $\D$ however it can be observed that the instability zone extends to larger dopings in the $\a=0$ case compared to the rotationally-invariant $\a=1$ case, as it was already observed for the models without crystal-field splitting in the above Sec.~\ref{sec:Kanamori_vs_Ising} and in Ref.~\onlinecite{demedici_el_comp}.

\begin{figure}[h!]
\begin{center}
\includegraphics[width=0.5\textwidth]{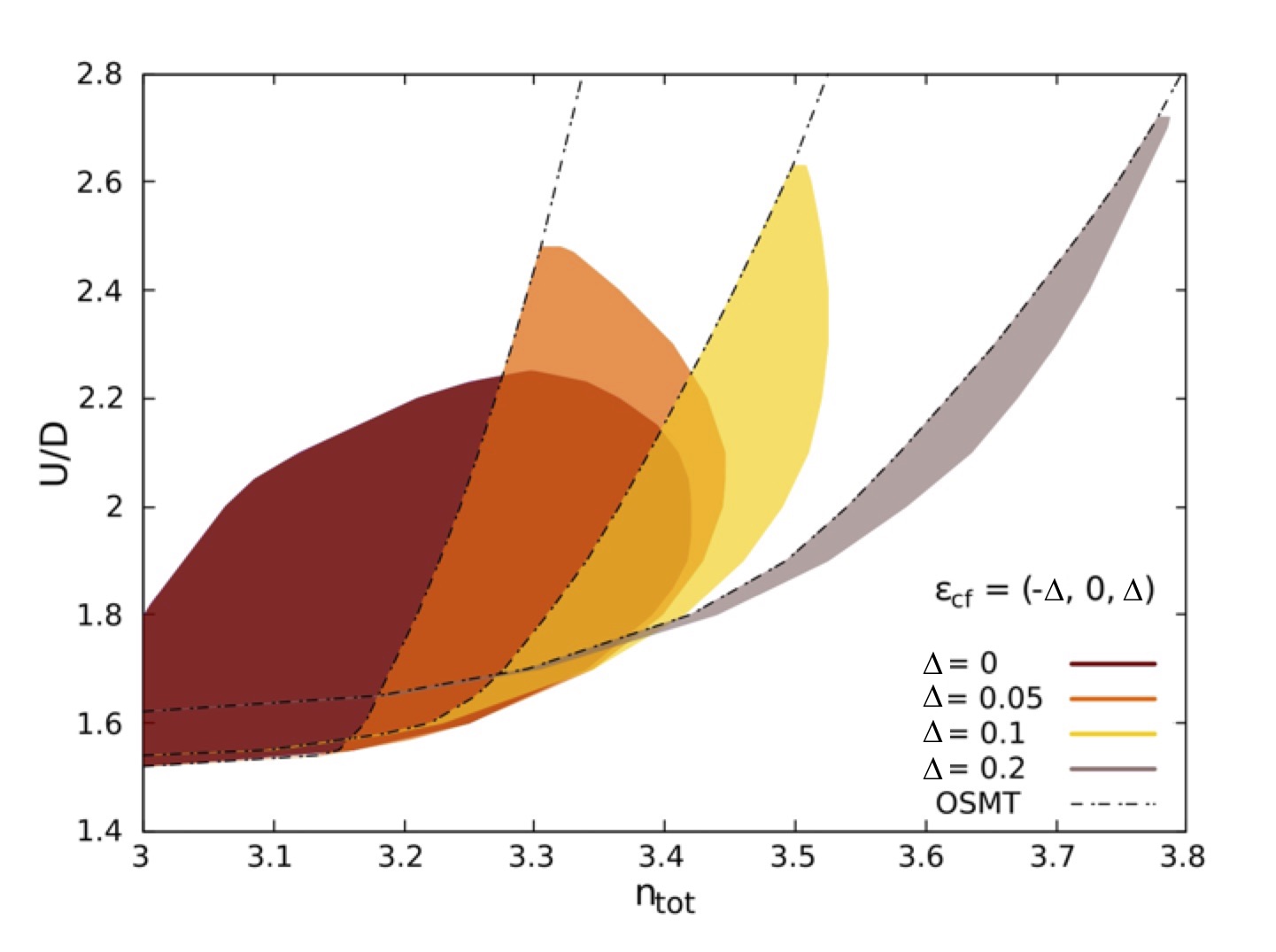}
\caption{Same as Fig.~\ref{fig:2b_cfs_SSMF} but for the 3-band model with symmetric crystal-field splitting among the bands $\D=\eps_1-\eps_2=\eps_2-\eps_3$ for density-density interaction ($\a=0$, $J/U=0.25$) within Slave-Spin Mean-Field.}
\label{fig:3b_cfs_SSMF}
\end{center}
\end{figure}

The same trends are seen for models with a larger number of orbitals. For $M>2$ one could consider different crystal-field splittings separating the various orbitals. In Fig. \ref{fig:3b_cfs_SSMF} the case $M=3$ for a symmetric splitting $\D=\eps_1-\eps_2=\eps_2-\eps_3$ is reported. It is found rather similar to the 2-orbital case, however it is clear that, albeit narrowed in the $U$ direction, the instability zone extends to a much higher doping than in the $\D=0$ case, the maximum doping reached being nearly doubled for $\D=0.2$.

\begin{figure}[h!]
\begin{center}
\includegraphics[width=0.475\textwidth]{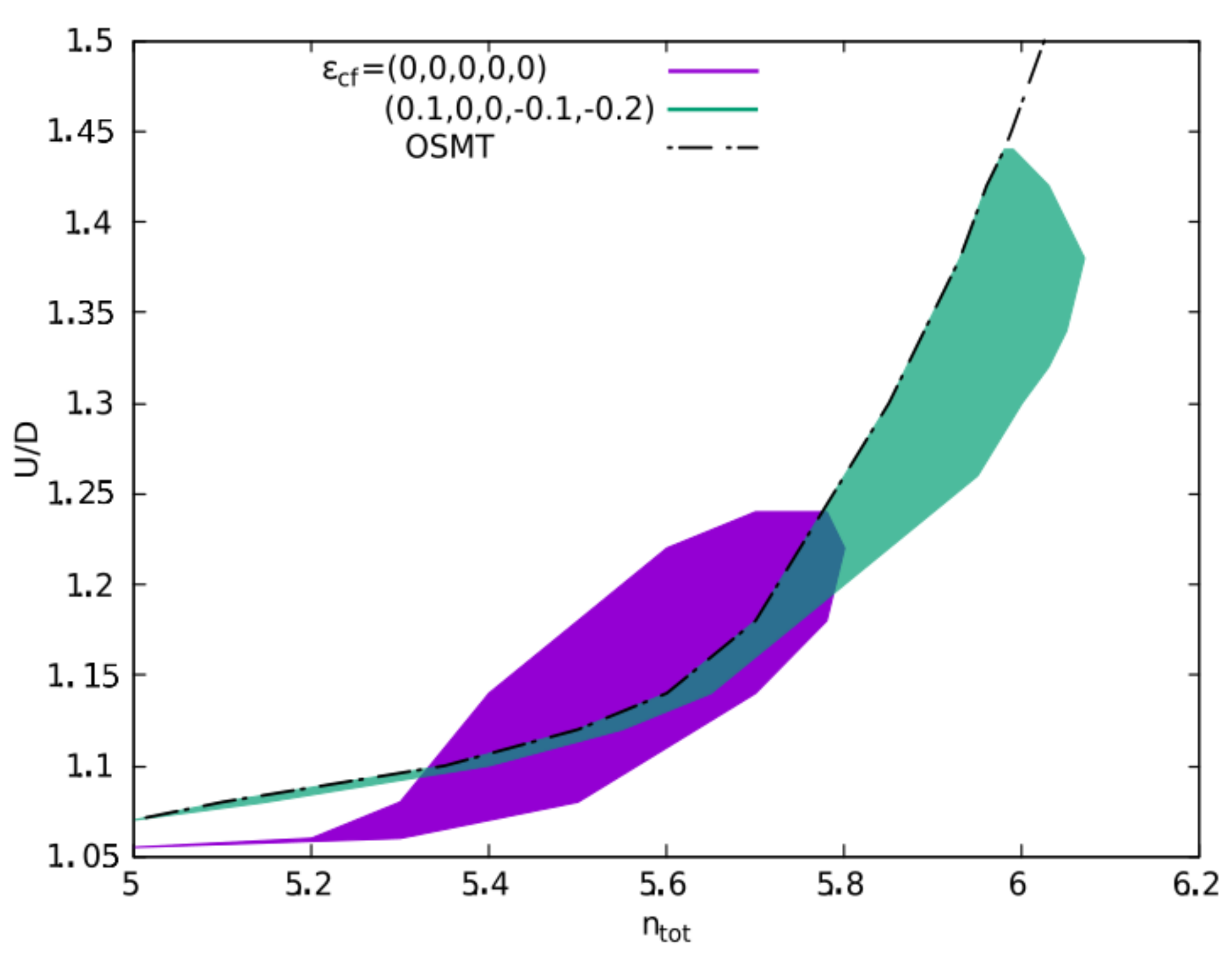}
\caption{Same as Fig.~\ref{fig:2b_cfs_SSMF} but for the 5-band model with crystal-field splitting among the bands typical of a tetragonal environment like that of iron-based superconductors, i.e., $\eps_1>\eps_2=\eps_3>\eps_4>\eps_5$ with $\D=\eps_1-\eps_2=\eps_2-\eps_4=\eps_4-\eps_5$, here calculated for density-density interaction ($\a=0$, $J/U=0.25$) within Slave-Spin Mean-Field.}
\label{fig:5b_cfs_SSMF}
\end{center}
\end{figure}
In Fig.~\ref{fig:5b_cfs_SSMF} we report a typical result for the $M=5$ case. Since the 5-orbital model is relevant for the FeSC, here we display the result for the case of a tetragonal symmetry, among the many possible crystal-field splitting cases, in which the two $e_{g}$ orbitals are split in energy and well below the $t_{2g}$ ones. Among these, two orbitals corresponding to the out-of-plane $t_{2g}$ remain degenerate, whereas one orbital is lifted in energy.
Strikingly, in this 5-orbital case the extension in doping of the instability zone to phase separation is very strongly enhanced, and even comparatively less "chopped" by the OSMT. 

\begin{figure*}
\begin{center}
\includegraphics[width=\textwidth]{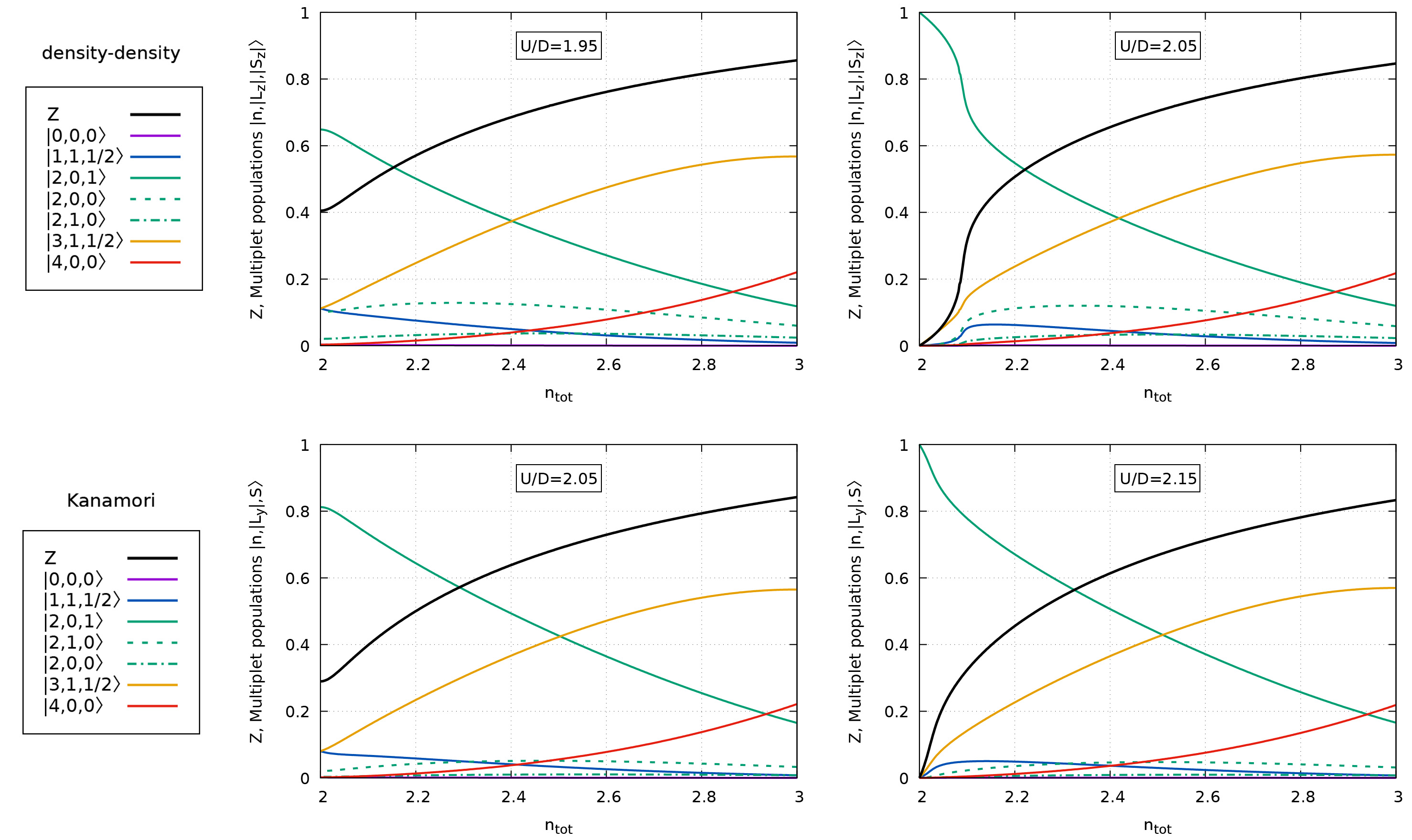}
\caption{Quasiparticle weight $Z$ (black line) and atomic multiplet populations (atomic-state amplitudes in the ground state times multiplet degeneracy - color lines) vs density for a degenerate 2-orbital Hubbard model for density-density interaction $\a=0$ (upper panels) and Kanamori $\a=1$ interaction (lower panels) for $J/U=0.25$ and $U$ values near the Mott transition ($U_c$). Left panels for $U<U_c$ ($Z$ is finite at half-filling), right panels for $U>U_c$ ($Z$ vanishes at half-filling).}
\label{fig:probs}
\end{center}
\end{figure*}

We stress the importance of this result for real materials: the instability zone cuts a wider range of dopings the larger the number $M$ of orbitals involved in the formation of the conduction bands, and the larger the crystal-field splitting between these orbitals.

We have verified that similar results are obtained with various crystal-field splitting configurations, which we do not show not to overweight the presentation.

\section{Interpretation in terms of local fluctuations and many body "extra" kinetic energy}\label{sec:interpret}

In this section we discuss a unifying principle to understand all the results we have discussed in the previous sections and the general tendency towards the charge instabilities of multi-orbital Hubbard models.
In order to have a more physical insight into these trends it is useful to look at the probability of occurrence of the possible local configurations. 

As mentioned in Sec.~\ref{sec:theory} , in the non-interacting half-filled system, in absence of crystal-field splitting, all configurations are realized with the same probability, and e.g. they appear with the same weight in the ground state wave function. The increase of the interaction $U$ reduces the weight of the configurations with filling different from half, until the Mott transition happens. In slave-particle mean-field approximations this weight is represented by the ground state amplitudes of the slave-particle variables corresponding to local configurations.  In the same approximations the weight of the configurations with filling different from half actually vanishes at the Mott transition and in the Mott state.

If we include a finite Hund's coupling keeping a fixed $J/U$ ratio, also the low-spin configurations are gradually eliminated and only those with the maximum possible spin survive in the Mott insulator. This is illustrated in Fig.~\ref{fig:probs} for the $M=2$ case: for both the density-density $\a=0$ and the Kanamori $\a=1$ interaction, at half-filling ($n=2$) the low-spin configurations are considerably suppressed with increasing $U$ and when $U>U_{c}$ the only configurations having a finite weight are the high-spin ones, i.e., those with total spin $S=1$ for the Kanamori case $\a=1$ and those with $|S_z|$=1 for the density-density case $\a=0$.\cite{Lechermann_RISB,FacioCornaglia-Mott_1st_2nd_Order,Fanfarillo_Hund,Haule_pnictides_NJP}

When we dope the system a metallic behavior is restored, and this is associated to a recovery of the lower-spin configurations at the expenses of the high-spin ones  and  an increase of charge fluctuations, i.e., an increased weight of configurations with a filling different from the average\footnote{For the doped system, a large value of $J$ - which can be in some cases near $U/3$ - can help these fluctuations further, since the atomic charge states with the different $n\geq M+1$ (and equivalently those for $n\leq M-1$) can differ in energy by as low as $U-3J$\cite{demedici_MottHund,Isidori-mixed_valence_Hund}.}. 
In Fig.~\ref{fig:probs} we compare the evolution of the weights as a function of doping for an interaction strength just below $U_{c}$ (left panels) and one just above $U_{c}$ (right panels). It is apparent that the solutions for $U<U_{c}$ and for $U>U_{c}$ differ significantly only below some specific doping and they become very similar in the large-doping region.

This doping value marks the crossover between a Hund's metal, which we identify with a doped high-spin Mott insulator, and a normal metal. This crossover happens at lower and lower doping and is more and more abrupt the closer $U$ is to $U_{c}$. A direct consequence of the quick revival of the fluctuations frozen by $U$ and $J$ is a correspondingly quick increase of the quasiparticle weight $Z$ at the crossover, as visible in Fig.~\ref{fig:probs}. 

At the lowest dopings the boundary of the Hund's metal coincides with the frontier of the instability zone towards phase separation reported in all the previous phase diagrams\cite{demedici_el_comp,demedici_Hunds_metals}. The insight that we are getting from this analysis is that the \emph{onset of the instability} at a given interaction strength $U$ \emph{is related to the rapidity of the crossover between the normal and the Hund's metal}.

We can strengthen this connection by observing that the quick revival of the fluctuations frozen by $U$ and $J$ leads to recover the "extra" multi-orbital kinetic energy that we discussed in Sec. II. This is explicitly illustrated within SSMF and for the half-filled system in Appendix B. The arguments used there are however general, and hold also for the doped case of interest here, albeit the analytic treatment becomes then more involved.
The computed kinetic energy (per site) $E_{kin}=\langle H_0\rangle/{\cal N}_{sites}$, where $\langle \rangle$ indicates the quantum average over the ground state - is zero in the Mott insulator within the slave-particle mean-fields. Doping leads to a negative value which increases in absolute value with the progressive delocalization of the carriers. Remarkably, it acquires a negative curvature as a function of doping
$\partial^2 E_{kin}/\partial n^2<0$.  
This has an important consequence, because the curvature of $E_{tot}$ is the inverse compressibility of the electronic system:
\be
\kappa_{el}^{-1}=\frac{\partial^2 E_{tot}}{\partial n^2}=\frac{\partial^2 E_{kin}}{\partial n^2}+\frac{\partial^2 E_{pot}}{\partial n^2}.
\ee
and, obviously, $E_{tot}=E_{kin}+E_{pot}$, where $E_{pot}=\langle H_{int}\rangle/{\cal N}_{sites}$.

\begin{figure}[h!]
\begin{center}
\includegraphics[width=0.48\textwidth]{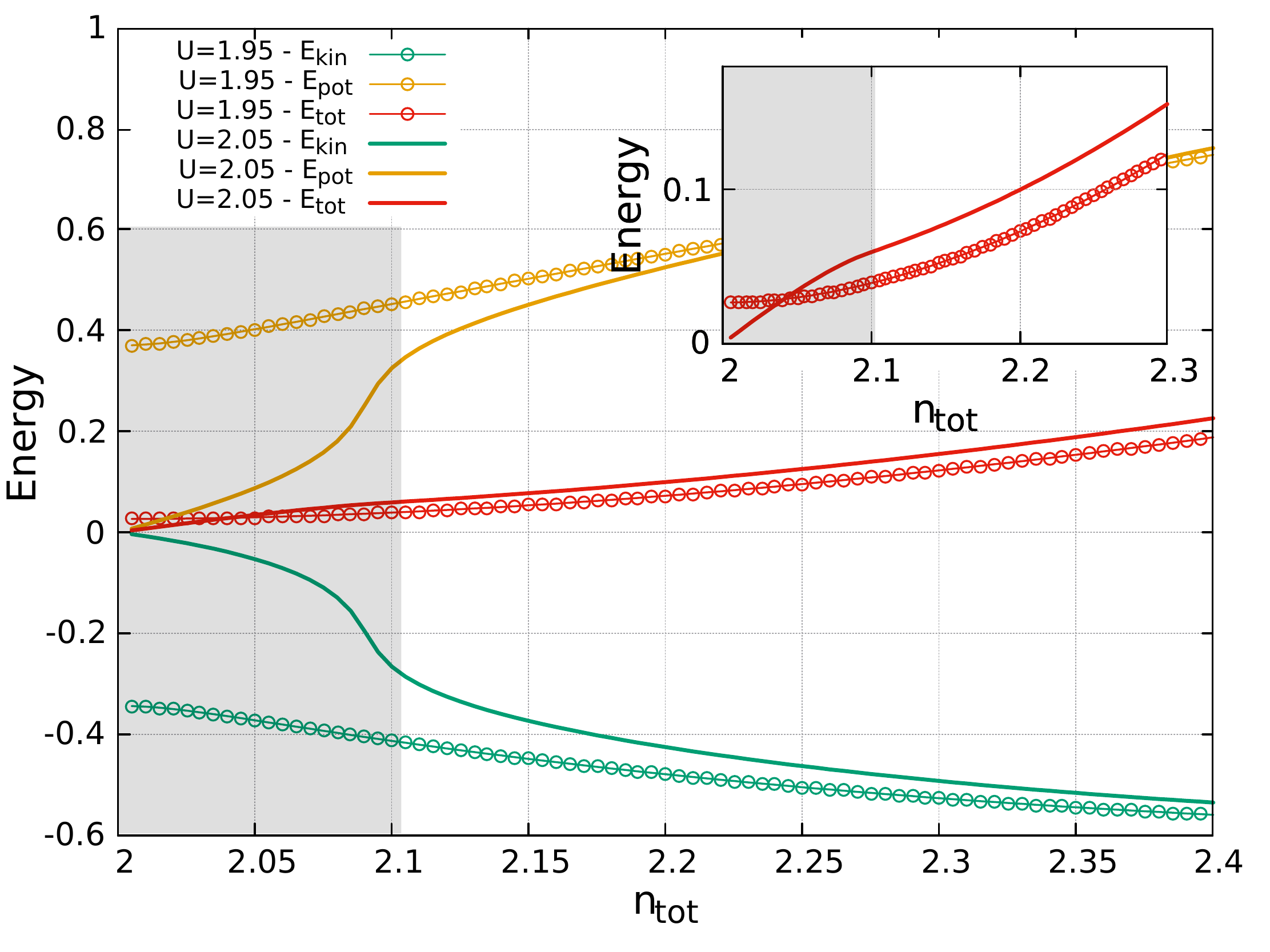}
\caption{Total, kinetic and potential energies per site for the 2-orbital degenerate Hubbard model with $J/U=0.25$, $\a=0$, for $U=1.95<U_c$($n=2$) and $U=2.05>U_c$($n=2$) as a function of the total density. The gray zone indicates the zone of phase separation for $U=2.05$. Inset: blow-up of the total energies highlighting the negative curvature of $E_{tot}$ for $U=2.05$ at low doping, corresponding to a negative compressibility.}
\label{fig:energies}
\end{center}
\end{figure}

The total energy $E_{tot}$ and the two contributions  $E_{kin}$ and $E_{pot}$ are all plotted as a function of the total density in Fig.~\ref{fig:energies} for the density-density $\a=0$ model, for the same cases $U<U_{c}$ and for $U>U_{c}$ of Fig.~\ref{fig:probs} (the arbitrary zero of energy is chosen such that the potential energy of the half-filled Mott insulator is zero). It is clear that despite the total energies are quite close in value, owing to the proximity in the phase diagram, the behavior of $E_{kin}$ and $E_{pot}$ highlights a substantial difference between the two metals found below and above the critical interaction strength. Indeed for the doped Mott insulator for $U=2.05D>U_{c}$ the crossover between the normal and the Hund's metal is signalled by the sharp change of behaviour in $E_{kin}$ and $E_{pot}$ around $n=2.08$. The crucial observation is that the negative curvature of $E_{kin}$ is not completely compensated by the positive curvature of $E_{pot}$, and results in an overall negative curvature of $E_{tot}$, and thus of a zone of negative compressibility (highlighted by the grey area, and zoomed in, in the inset).  On the other hand the metal below $U_{c}$ does not show this behavior.
 
Similar results are found for any form of interaction, crystal-field splitting and number of orbitals, shedding light on the nature of the phase separation instability highlighted by the calculated diverging/negative compressibility, and its universal presence for Hund's correlated systems.

Moreover, this interpretation also explains the various trends reported in this article. Our first main result is that the instability is more pronounced and more extended in the case of density-density interaction, compared to the case of rotational-invariant Kanamori interaction. Indeed the quenching of multi-orbital fluctuations is more radical in the density-density case: only the doublet with $|S_z|=S_{max}(n)$ is left degenerate at low energy among the local configurations in the former, whereas rotational invariance preserves the degeneracy of the whole low-energy multiplet with $S^2=S_{max}(S_{max}+1)$ in the latter. 
Therefore in the $\a=0$ case the quenching of the extra multi-orbital kinetic energy is stronger, and its doping-driven release is more abrupt, causing a stronger inversion of curvature, and in turn a  larger instability zone. 

A similar interpretation can be given to the fact that a crystal-field splitting can push the instability zone to larger doping.
Indeed we have shown that doping a high-spin Mott insulator results in an OSMP at low doping, because only a subset of orbitals is actually doped and contributes to the metallic behavior, while the rest of the system remains insulating. 
In the 2-orbital case, as long as we are in the OSMP, only the  orbital lowered in energy by the crystal field is doped. This implies that the multi-orbital "extra" kinetic energy due to a quantum coherent superposition of configurations with different orbital populations cannot be activated, since one orbital is still quenched in its singly occupied state. It is only after both orbitals are doped and charge fluctuates in both of them that the multi-orbital fluctuations among all configurations are restored, and hence the extra kinetic energy as well.

Therefore, in a multi-orbital system in general, the phase separation instability due to the presented mechanism can happen as soon as at least two orbitals are doped, and thus in cases where multiple OSMP zones (with an increasing number of metallic orbitals) are present in the phase diagram an unstable zone can be found nearby each OSMT (on the more metallic side of it). In particular, as we highlight in Figs. \ref{fig:3b_cfs_SSMF}  and \ref{fig:5b_cfs_SSMF}, the instability zone is still found in the fully metallic phase, in the vicinity of the outermost OSMT, the one that happens at the largest doping and at which the orbital(s) at highest energy become localized. This is the ultimate reason why the instability region in the presence of crystal-field splitting shifts to larger dopings\footnote{It is worth mentioning that we have found (not shown) also further instability zones within the OSMP, close to each successive OSMT}.

\section{Application to FeSe bulk vs Monolayer}\label{sec:FeSe}

We now discuss the relevance of the present analysis in iron-based superconductors as an example for the interest of our results in the theoretical analysis of material-specific properties of actual strongly-correlated solids.

In particular we interpret some of the results on bulk FeSe and its monolayer form reported by two of the authors in Ref.~\onlinecite{VillarArribi_FeSe_el_comp}. 
There, a realistic simulation of these two materials was performed based on a density-functional theory band structure calculation, parameterized with maximally-localized Wannier functions, yielding a material-specific $H_{0}$ for each one of the two cases, namely bulk and monolayer. These 5-orbital tight-binding bare Hamiltonians were supplemented by $H_{int}$ of the form (\ref{eq:ham_kanamori}) (and $\a=0$) and solved within SSMF.

The compressibility was calculated and shown to be enhanced around, diverging at the border, and negative inside, of a moustache-shaped zone departing from half-filling at $U_{c}$ and extending into the $U$-doping plane, following the general trends found in Hund's metals and in SSMF simulations of FeSC\cite{demedici_el_comp,demedici_Hunds_metals}.

\begin{figure}[h!]
\begin{center}
\includegraphics[width=0.48\textwidth]{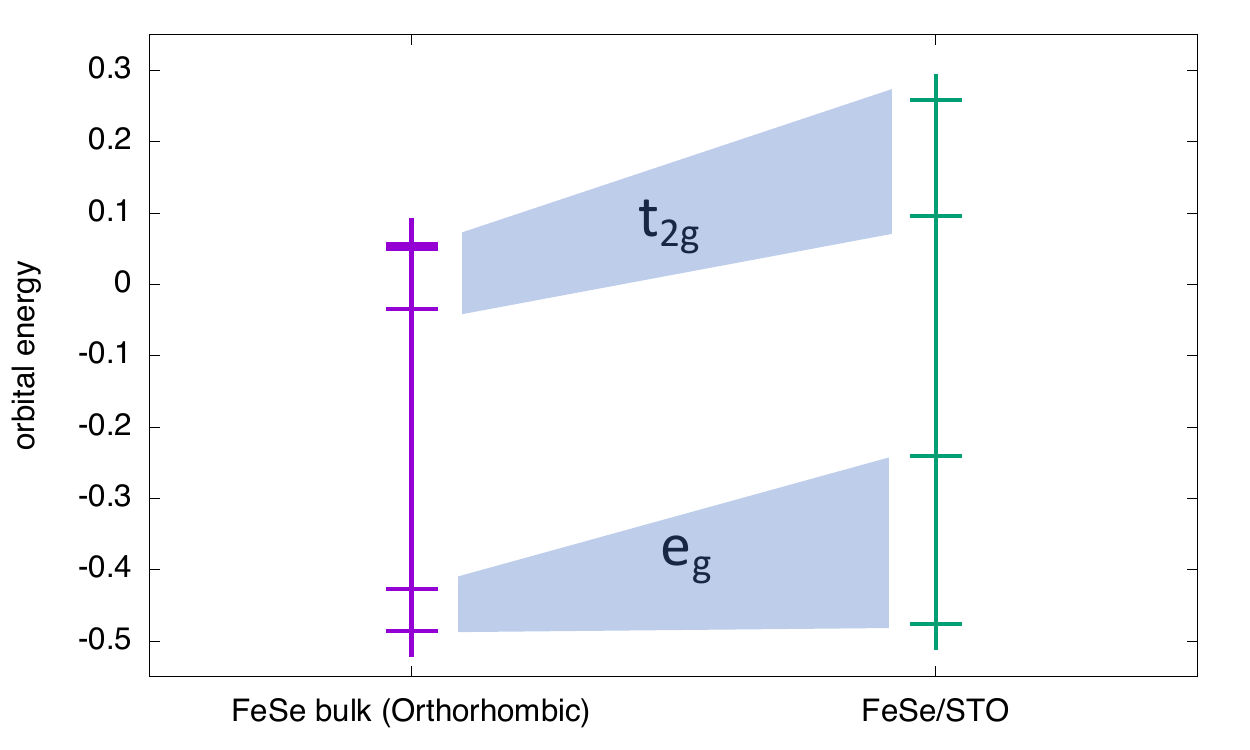}
\caption{Values of the orbital energies in the tight-binding parametrizations of the DFT band structure for bulk and monolayer FeSe calculated in Ref.~\onlinecite{VillarArribi_FeSe_el_comp}. The overall crystal-field splitting between the lowermost in energy $e_g$ orbital and the uppermost $t_{2g}$ one is larger ($\sim$0.7eV) in the monolayer case compared to the bulk ($\sim$0.5eV), motivating a larger extension of the instability zone found in the realistic simulations\cite{VillarArribi_FeSe_el_comp}, along the insight given by the models analyzed in the present work.}
\label{fig:cfs_FeSe_mono_bulk}
\end{center}
\end{figure}

Interestingly, the instability zone was found more pronounced and extended in FeSe monolayer than in the case of bulk FeSe, which can be seen as an indication that the enhancement of compressibility plays a role in obtaining the record high-$T_{c}$ superconductivity reported for monolayer FeSe\cite{Ge_FeSe_mono_Tc100}.

Following our analysis of the effect of a crystal-field splitting on the doping extension of the instability zone in Sec. ~\ref{sec:cfs} it is natural to conclude that the larger instability zone of the monolayer simulations is due to the enhanced crystal-field splitting of the corresponding $H_{0}$. The values of orbital energies obtained in the tight-binding parametrization are reported in the figure Fig. \ref{fig:cfs_FeSe_mono_bulk}.
It is worth mentioning that, in these simulations, no actual OSMT happens at zero-temperature that would allow to use slavishly the arguments given here for the simplified models. But in the realistic models, even if the inter-orbital hopping prevents a strict OSMT from happening, there is still a clear frontier between a non-selective metal and an orbital-selective one, where some orbital(s) (for the FeSC, the $d_{xy}$ orbital) become  extremely correlated, and in some cases almost localized\cite{demedici_OSM_FeSC,Yi_Universal_OSM_Chalcogenides,Hardy_KFe2As2_Heavy_Fermion}. The extremely reduced charge fluctuations in these almost localized orbitals, bring about the reduced weight in the configurations responsible for the enhanced kinetic energy, and thus allow our analysis to apply.

\section{Conclusions}\label{sec:conclusions}

We have analyzed several multi-orbital Hubbard models (with $M=2,3,5$ orbitals) in presence of a finite Hund's coupling, focusing in particular on the occurrence and extension of a charge-instability zone of the Hund's metal phase, previously found both in models\cite{demedici_el_comp} and realistic simulations of Fe-based superconductors\cite{VillarArribi_FeSe_el_comp} and related materials\cite{Edelmann_Chromium_analogs}.  This instability occurs universally in Hund's metal simulations and is signaled by a divergent/negative electronic compressibility for a zone of the U-density phase diagram, where the normal paramagnetic homogeneous metal cannot exist and is thus unstable towards phase separation. The unstable zone is surrounded by an area where the compressibility is enhanced, which can favour other instabilities including superconductivity. 

We have identified two general trends which are useful to understand and motivate realistic calculations for specific correlated materials:
The phase separation region always originates from the Mott transition at half filling and
\begin{itemize}
\item the doping range spanned by the instability zone increases with the number of orbitals $M$ and generically with the crystal-field splitting of the local energies of the orbitals,
\item the spin-anisotropy (density-density form) of the local interaction enhances the instability compared to a perfectly symmetric rotational-invariant Kanamori interaction
\end{itemize}

These trends are interpreted in terms of the extra kinetic energy associated with the multi-orbital fluctuations, which are suppressed by Hund's coupling around the half-filled Mott insulator, and suddenly restored by the doping at the Hund metal to normal metal frontier, thereby driving the instability. Both the breaking of the spin-rotational (by the density-desnsity form of the interaction) and of the orbital-rotational invariance accentuate this quenching, and the corresponding release with doping, thus accentuating the ensuing charge instability.

The first one of the above points in particular helps explaining the considerable prominence of the instability in the monolayer FeSe/STO simulation, compared to the one of bulk orthorhombic FeSe, which had already been put in correlation with the striking difference of the experimentally reported T$_c$ (8K in the bulk, $\gtrsim$ 65K in the monolayer)\cite{VillarArribi_FeSe_el_comp}.

\appendix
\section{Slave-spin mean-field method}\label{appendix:SSMF}

Here we briefly derive the slave-spin mean-field equations, specialized to the case explored in this work of identical bands without inter-orbital hopping and in presence of a crystal field.

A multi-orbital Hubbard model with Hamiltonian $\hat H$ eqs. (\ref{eq:H0})+(\ref{eq:ham_kanamori}) can be rigorously expressed on an enlarged Hilbert space, provided averages are restricted by a constraint. 

Indeed at each site $i$ each one of the original fermions (created by the operator $d_{im\s}^\+$ - labeled by orbital and spin indices $m\s$) has two states, $|0\rangle$ and $|1\rangle$. An auxiliary system can be considered, that has at each site and for each orbital/spin flavor $m\s$ both a pseudofermion (created by $f_{im\s}^\+$) and a spin-$1/2$  (flipped by $S^\pm_{im\s}$ and of which the $z$ component is measured by $S^z_{im\s}$). The auxiliary system will have a local Hilbert space for each $m,\s$ spanned by the four states  $|0\rangle_f |0\rangle_s$, $|1\rangle_f |1\rangle_s$, $|0\rangle_f |1\rangle_s$  , $|1\rangle_f |0\rangle_s$ (where the subscripts f and s indicate the pseudofermion and the spin respectively, and for the spins we use the notation "0" for the "down" state and "1" for the "up" state), which is the direct product of the respective local Hilbert spaces of the two auxiliary variables.
The first two of these four states (dubbed the "physical" states of the auxiliary space, for which the condition \be
n^f_{im\s}=S^z_{im\s}+1/2,
\label{eq:SS_constraint}
\ee 
holds, with $n^f_{im\s}\equiv f^+_{im\s}f_{im\s}$) are associated to the states $|0\rangle_d$ and $|1\rangle_d$, respectively, of the original fermion. 

An exact mapping of the original problem is obtained if one can define a Hamiltonian operator, acting on the auxiliary Hilbert space, which has the same matrix elements among the physical states as the original Hamiltonian, and a constraint excluding from all quantum or ensemble averages all states other than the "physical" ones. 

In practice, this is not a simplification of the original problem. However, one can make approximations on the auxiliary problem which in the end are less drastic than those performed on the original one. Here we follow the original formulations of a mean-field approximation introduced in Refs. \onlinecite{demedici_Slave-spins} and \onlinecite{Hassan_CSSMF}, and extensively discussed in Ref. \onlinecite{demedici_Vietri}. 

The interaction part of the Hamiltonian eq. (\ref{eq:ham_kanamori}), for the density-density-only case ($\a=0$) can easily be expressed in terms of the spin variables only:
\bea
H_{int}^s&&=U\sum_{im}S^z_{im\up}S^z_{im\down} 
+ U' \sum_{im\neq m'}S^z_{im\up}S^z_{im'\down} \nonumber\\
&&+ (U'-J) \sum_{im<m'\s}S^z_{im\s}S^z_{im'\s}\label{eq:Hsint}
\eea
The one-body part instead involves both slave-spins and pseudofermions:
\be
H_0=\sum_{i\neq jmm'\s}t_{ij}^{mm'} f_{im\s}^\+f_{jm'\s}O_{im\s}^\+O_{jm'\s}+\sum_{im\s}\eps_{m}f_{im\s}^\+f_{im\s}\label{eq:H0_SS}
\ee
where $O_{im\s}^\+\equiv S^\+_{im\s}+c^*_{im\s}S_{im\s}^-$ is an adapted form of the raising spin operator. If the constraint eq. (\ref{eq:SS_constraint}) is enforced exactly, then the operator $f_{im\s}^\+O_{im\s}^\+$ is an exact mapping of $d_{im\s}^\+$ in the enlarged Hilbert space, and the value of the parameter $c_{im\s}$ is immaterial. In any approximation scheme in which the constraint cannot be exactly enforced,  the choice for $c_{im\s}$ matters in the action of $f_{im\s}^\+O_{im\s}^\+$ on the unphysical states. But this gauge can be used at our advantage as shown later in this section.

In our scheme three approximations are performed. First, we mean-field decouple slave-spin and pseudofermion operators in the hopping term thus leaving us with a free-fermion Hamiltonian (since the interaction is entirely treated by the slave-spin variables), and a lattice spin model where several slave-spins interact on site and are also coupled to the slave-spins of neighboring sites. Second, we perform a mean-field decoupling of the latter lattice system in the spirit of Weiss mean-field, and thus we are left with a single-site Hamiltonian in an effective field. Third, we treat the constraint with site-independent (and spin-independent, since we address the paramagnetic phase here) Lagrange multipliers $\l_m$. Details of this procedure can be found e.g. in Ref. \onlinecite{demedici_Vietri}.

The resulting mean-field Hamiltonian (specialized to the case treated in this work, with intra-orbital hopping only, equal for all bands, i.e., $t_{ij}^{mm'}=t_{ij}\d_{mm'}$) is $H-\mu N=H_s+H_f-\mu N$, with $H_f-\mu N$ in eq. (\ref{eq:QP_ham}) and  $H_s= \sum_i H_s^i$ with:
\begin{align}
&H_s^i=\sum_{m\s} (h_m O_{m\s}^\++H.c.)+\l_m (S_{m\s}^z+\frac{1}{2})+H_{int}^s[i],\\
&\langle n^f_{m\s}\rangle=\langle S^z_{m\s}\rangle + \frac{1}{2}, \qquad \forall \, m,\s,
\end{align}
where the self-consistent parameters are:
\begin{align}
&Z_m= |\langle O_{m\s} \rangle |^2, \label{eq:Z_O} \\
&h_m= \langle O_{m\s} \rangle\sum_{j\neq i} t_{ij}\langle f_{im\s}^\+ f_{jm\s}\rangle=\langle O_{m\s} \rangle \eps_0(n_{m\s}),\label{eq:hm_O}
\end{align}
and 
\be
\eps_0(n_{m\s})\equiv \sum_k \eps_k \langle f_{km\s}^\+ f_{km\s}\rangle=\int_{-\infty}^\mu \!\!d\eps \, (\eps-\eps_m)\, D(\eps),
\ee
is the bare band kinetic energy at $T=0$ ($\eps_k$ and $D(\eps)$ are the band dispersion and the density of states respectively, as defined in Section \ref{sec:models_methods} of the main text).The $n_{m\s}=\langle f_{m\s}^\+f_{m\s}\rangle=\langle d_{m\s}^\+d_{m\s}\rangle$ are the orbital populations, which can still differ, along with $h_m$, $\l_m$, etc. because of the possible different values of $\eps_m$.

Here the gauge freedom represented by the choice of the $c_{im\s}$ can be used for this mean-field approximation to reproduce known limits of the model. In particular, we choose here the value - real and independent of the site -  that correctly yields $Z_{m}=1$ in the non-interacting limit $U=J=0$. This is  
\be\label{eq:cm}
c_{i m\s}=\frac{1}{\sqrt{n_{m\s}(1-n_{m\s})}}-1\equiv c_m.
\ee
This expression is extended to finite values of the interaction and evaluated self-consistently (i.e., using the interacting orbital populations).
The real value chosen for the $c_{im\s}$ entails the reality of the Hamiltonian, which implies that $\langle S^+_{m\s}\rangle=\langle S^-_{m\s}\rangle=\langle S^x_{m\s}\rangle$ and thus eqs. (\ref{eq:Z_O}) and (\ref{eq:hm_O}) acquire the more insightful form:
\begin{align}
&Z_m= (1+c_{m})^2\langle S^x_{m\s}\rangle^2, \label{eq:Z_Sx} \\
&h_m= \sqrt{Z_m} \eps_0(n_{m\s}).  \label{eq:hm_Sx} 
\end{align}
This illustrates that the metallic state $Z_m\neq0$ is signalled by a non-zero magnetization of the slave-spin lattice in the $x$ direction induced by the self-consistent field $h_m O_{m\s}^+ + h_m^* O_{m\s}=(1+c_m)h_m S_{m\s}^x$ which also points in the x direction ($h_m$ is real and negative), while the Mott insulating state is the corresponding disordered phase in the same direction.

In the same non-interacting limit one should also get $\l_m=0$, so that the quasiparticle dispersion coincides with the original dispersion of the non-interacting electrons. However the present mean-field formulation yields finite $\l_m(U=J=0)=\frac{(2n_{m\s}-1)}{n_{m\s}(1-n_{m\s})}h_{m\s}$. We correct this unwanted feature by altering the bare orbital energies so to compensate exactly for this artificial shift and reproduce the correct orbital populations in the non-interacting limit. In other words, we evaluate the above expression at $U=J=0$ defining $\l_m^0\equiv \l_m(U=J=0)=\frac{(2n_{m\s}-1)}{n_{m\s}(1-n_{m\s})}h_{m\s}$ and add a term $+\l^0_mf_{im\s}^\+ f_{im\s}$ to the quasiparticle hamiltonian eq. (\ref{eq:QP_ham}).

For a slightly different formulation leading to very similar mean-field results see Ref. \onlinecite{YuSi_LDA-SlaveSpins_LaFeAsO}. The final mean-field equations of the two formulations are identical, except for the formula used for the shift $\l_m^0$, which for this alternative formulation reads $\l^0_m=\sqrt{Z_m}\frac{(2n_{m\s}-1)}{n_{m\s}(1-n_{m\s})}h_{m\s}$. It indeed coincides with ours in the non-interacting limit (where $Z_{m}=1$), but importantly it is evaluated self-consistently at any $U$ and $J$ which gives improved results\cite{Pizarro_thesis}.

\section{Expansion of the slave-spin equations: enhancement of the kinetic energy with the ground state degeneracy in proximity of a Mott insulator}\label{appendix:KinEn_exp}

In the slave-spin mean-field approximation a Mott insulator is a solution in which $Z_{m}=0$. By the self-consistency equation eq. (\ref{eq:hm_Sx}), this implies $h_m=0$.
Thus in proximity of a Mott insulator a perturbative treatment in $h_m$ of the slave-spin problem can be performed. 

Indeed, the actual value of $h_m$ for a given interaction strength and filling is determined by the self-consistency equation eq. (\ref{eq:hm_Sx}) for which (in absence of inter-orbital hopping) $h_m$ is a linear function of $\langle S^x_{m\s}\rangle$. The latter is calculated from the spin Hamiltonian and is instead a more complicated function of the $h_m$'s, that we can calculate explicitly using perturbation theory. At linear order in $h_m$ the only solution of these equations is trivially in $h_m=0$, so that higher orders are needed to determine a non-trivial solution yielding a finite $h_m$. However, for small enough $h_m$ (i.e. close enough to a Mott insulator) the linear term will anyway dominate (and also ultimately determine the critical parameters for the Mott transition\cite{demedici_Vietri}). 

Here we show that the kinetic energy at leading order in $h_m$ is indeed enhanced by the ground state degeneracy. The actual kinetic energy of the system should be evaluated at the self-consistent value of $h_m$, but the prefactor to the  $h_m^2$ term that we calculate here is enough to illustrate this dependence when $h_m$ is small enough.

Let's restrict ourselves to the half-filled, particle-hole symmetric case with no crystal-field splitting (i.e., $\eps_m=0$), for which $\mu=\l_m=0$ and $c_m=1$. 
Then the unperturbed Hamiltonian is simply $H_{int}^s[i]$ eq. (\ref{eq:Hsint}). 
At $T=0$ we only need the ground state of this Hamiltonian, which will have a different degeneracy $d_0$ depending on the value of $J$, so later on we will distinguish the $J=0$ from the finite $J$ case.

The perturbing Hamiltonian is $V\equiv H_s^i-H^s_{int}[i]=2h\sum_{m\s}2S^x_{m\s}=2h\sum_{m\s}(S^+_{m\s}+S^-_{m\s})$, where $h\equiv h_m$ is here equal for all orbitals. 

It simply flips any of the slave-spins. 
It removes the ground state degeneracy at the second order in $h$ (the perturbation has no matrix elements within the degenerate subspace), and in order to obtain the "correct" unperturbed ground state $|\phi_0^{(0)}\rangle$ (the one to which the perturbed ground state tends for $h\rightarrow 0$) one has to diagonalize the matrix $H'\equiv V(E_0-H_{int}^s)^{-1}V$ in the degenerate subspace, where $E_0$ is the unperturbed ground state energy. The ground state ket will have a correction at the linear order instead, which according to standard perturbation theory reads: $|\phi_0^{(1)}\rangle=|\phi_0^{(0)}\rangle+ \sum_{|s\rangle\neq |\phi_0\rangle} \langle s|(E_0-H_{int}^s)^{-1}V|\phi_0^{(0)}\rangle |s\rangle$. 

The kinetic energy of the system per site in this mean-field approximation also happens to be the average value of the perturbation  $E_{kin}\equiv \langle H_0\rangle/{\cal N}_{sites}= \langle V\rangle$ (where $H_0$ is given in eq. (\ref{eq:H0_SS})). To leading order in perturbation theory it reads: 
\be
E_{kin}=\langle \phi_0^{(1)} |V |\phi_0^{(1)}\rangle=2\langle \phi_0^{(0)} |V(E_0-H_{int}^s)^{-1}V |\phi_0^{(0)}\rangle,\label{eq:Ekin_pert}
\ee
(where we used explicitly the fact that $\langle \phi_0^{(0)} |V|\phi_0^{(0)}\rangle=0$) which is also twice the ground state energy correction to leading order.

Eq. (\ref{eq:Ekin_pert}) illustrates that the kinetic energy is the number of ways in which flipping any two slave-spins brings the ground state into itself, weighed by $2/(E_0-E_S)$, twice the inverse of the (negative) energy difference between the ground state and the intermediate excited state. 

In physical terms this tracks the number of processes by which a particle can hop onto a neighboring site and back to any of the spin-orbitals still turning the ground state into itself. This is in strict analogy with the perturbative arguments determining Kondo coupling of an impurity in a bath\cite{schrieffer_wolff_physrev_1966,coqblin_schrieffer_1969}, which through Dynamical Mean-Field Theory describe the itinerancy of particles in a lattice model\cite{Georges_RMP_dmft}.

Now these processes are enhanced when the ground state has a greater degeneracy. We illustrate this in the following by comparing the case at $J=0$ with that for finite $J$.

\medskip

\begin{itemize}
\item \emph{J=0, SU(2M) symmetry}
\end{itemize}

At $J=0$, up to a constant shift, $H^s_{int}[i]=U/2(\sum_{m\s}S^z_{m\s})^2=U/2(S^z_{tot})^2$. The system has an even number $2M$ of slave-spins on each site, hence any state with $S^z_{tot}=0$ is a ground state.
Owing to the SU($2M$) symmetry of the $J=0$ problem, there are $d_0= {2M\choose M}$ such states ($| S^z\!=\!0\,; \,  l\, \rangle$, for $l=1\ldots d_0$) , corresponding to the number of ways to take half of the $2M$ spins up and half down. Physically this corresponds to the ways of putting $M$ particles in $2M$ spin-orbitals, owing to the half-filling of the system.

All the states with one flipped spin are $U/2$ higher in energy from the ground state and hence one can diagonalize $H'=-2V^2/U$. 
The lowest-energy eigenstate of the restriction of $H'$ to the unperturbed degenerate manifold is
\be
|\phi_0^{(0)}\rangle=\frac{1}{\sqrt{d_0}}\sum_{l=1}^{d_0} | S^z\!=\!0\,; \,  l\, \rangle
\ee
i.e., the linear combination of all the degenerate basis states in the ground state manifold with all plus signs. This can be easily checked by inspection.
Indeed $H'$ flips down any of the $M$ spins pointing up, and then flips up any of the now $M+1$ spins pointing down. The analogous process takes place starting with a flip up. This makes $2M(M+1)$ possible processes. The fact that \emph{all of the $d_0$} degenerate basis states with $M$ spins up and $M$ spins down are included in the linear combination ensures that all these "exchange" processes are active. The plus signs in the linear combination also ensures that all the corresponding $-U/2$ contributions add up, generating the lowest possible energy.
This, by inserting $|\phi_0^{(0)}\rangle$ into eq. (\ref{eq:Ekin_pert}) results in:
\be
E_{kin}=- \frac{32h^2}{U} M(M+1)
\ee

\medskip

\begin{itemize}
\item \emph{J$\neq$0, Density-Density interaction, $Z_2$ symmetry}
\end{itemize}

For $J\neq0$ and $M=2$, the excitation energy is $U_{eff}/2$ with $U_{eff}=U+(M-1)J$, so the matrix to be diagonalized to find the "correct" unperturbed ground state is the restriction of $H'=-2V^2/U_{eff}$. For higher $M$ the excited multiplets are split by $J$, and using $-2V^2/U_{eff}$ is an approximation, not qualitatively altering the present line of thoughts, however.

$H_{int}^s[i]$ now splits the manifold with $S^z=0$ and for the case of density-density only interaction ($\a=0$) the ground state is \emph{only two times degenerate}. The two degenerate states are  $|1,\ldots ,1,0,\ldots ,0\rangle$, with all $M$ slave spins corresponding to spin-orbitals $m\up$  - in our ket notation the first $M$ of all the $2M$ slave-spin - pointing "up" (1, in our notation) and the remaining ones pointing "down" (0 in our notation), or the inverse $|0,\ldots ,0,1,\ldots ,1\rangle$.

This degeneracy is split at order $h^{2M}$, but the "correct" unperturbed ground state is still the linear comibnation of these two basis states with the plus sign:
\be
|\phi_0^{(0)}\rangle=\frac{1}{\sqrt{2}}(|1,\ldots ,1,0,\ldots ,0\rangle+|0,\ldots ,0,1,\ldots ,1\rangle)
\ee

However to the leading order $h^2$ only $2M$ processes are active: those flipping twice a given slave spin. 
Thus inserting $|\phi_0^{(0)}\rangle$ into eq. (\ref{eq:Ekin_pert}) gives:
\be
E_{kin}=- \frac{32h^2}{U_{eff}} M=- \frac{32h^2}{U+(M-1)J}M
\ee
which for any $M\geq 2$ is much smaller than in the $J=0$ case.
This illustrates our point that the degeneracy of the ground state enhances the kinetic energy through the activation of extra hopping channels.

The treatment of the Kanamori Hamiltonian ($\a=1$) cannot be done to the same level of accuracy in the present slave-spin formulation but the degeneracy being $M+1$ thus still much smaller than in the $J=0$ case, the reduction of the Kinetic energy due to $J$ holds in that case too.
Again this parallels known results on the Kondo temperature of high-spin impurities\cite{schrieffer_japplphys_1967, Georges_annrev}.

It is worth mentioning that the dependence of the critical interaction strength for the Mott transition is a direct concretization of the kinetic energy dependence just shown. Indeed the value of 
\be\label{eq:2Sx_vs_h}
\sqrt{Z_m}=\langle 2 S^x_{m\s}\rangle=E_{kin}/(4hM)=
\begin{cases}
&- \frac{8h}{U} (M+1)\\
&- \frac{8h}{U+(M-1)J} \\
\end{cases}
\ee
respectively for $J=0$ and finite $J$,
can be inserted in the self-consistency equation eq.(\ref{eq:hm_Sx}) to extract the value of the interaction for which the non-trivial solution too reaches $h_m=0$. 
This yields 
\be
U_c=-8(M+1)\bar \eps,
\ee for $J=0$ and any value of $M$ (where we used the notation $\bar\eps\equiv\eps_0(n_{m\s}=1/2)\leq0$), and
\be 
U_c=-8\bar \eps-(M-1)J,
\ee
for finite $J$ and $M=2$ (exact) or larger (approximate).

\section{Mott gap edge and width of the Hubbard bands}\label{appendix:Hubbard_bands}

Here we show that the same mechanism leading to the enhancement of the quasiparticle kinetic energy detailed in the previous section leads to the enhancement of the distance of the Mott gap edge from the (lowest, if $J$ is finite) atomic excitation obtained adding or subtracting one particle, which can be interpreted as half the width of the Hubbard band. As mentioned in section \ref{sec:theory}, this converges with similar arguments used in Ref. \onlinecite{gunnarsson_multiorb}. 

Indeed if the chemical potential is moved inside the gap of the half-filled Mott insulator studied in appendix \ref{appendix:KinEn_exp} (in absence of crystal-field splitting, $\eps_m=0$) the particle symmetry is lost and $\l_m$ will be nonzero. However half-filling imposes $\l^0_m=0$ and through eq. (\ref{eq:QP_ham}) also that $\lambda_m=-\mu$. Eq. (\ref{eq:cm})) still gives $c_m=1$. 

For finite $\l_m$ eq. (\ref{eq:2Sx_vs_h}) becomes:

\be\label{eq:2Sx_vs_h_lam}
\sqrt{Z_m}=\langle 2 S^x_{m\s}\rangle=
\begin{cases}
&- \frac{8hU}{U^2-4\l_m^2} (M+1)\\
&- \frac{8h \, U_{eff}}{U_{eff}^2-4\l_m^2} \\
\end{cases}
\ee
respectively for $J=0$ and finite $J$.
Again inserting this relation in the self-consistency equation eq.(\ref{eq:hm_Sx}) one extracts the condition for which the non-trivial solution too reaches $h_m=0$. For each $U> U_c$ this gives a critical value for the chemical potential for which the density-driven Mott transition happens:
\be\label{eq:mu_c}
\mu_c=-\l_m=\pm \frac{1}{2}
\begin{cases}
&\sqrt{U^2+\bar\eps \, 8U (M+1}) \qquad (J=0),\\
&\sqrt{U_{eff}^2+\bar\eps \,8U_{eff}} \qquad \;\; \quad (J\neq 0). \\
\end{cases}
\ee

The distance $\D E_{Hub}$ between $\mu_c$ and the atomic excitation energy ($U/2$ and $U_{eff}/2$ respectively for $J=0$ and finite $J$)  around which the Hubbard band disperses can be interpreted as half the width if the Hubbard band. Expanding at large $U$ one gets a value for this distance which is independent of $U$ and reads 

 \be\label{eq:mu_c}
\D E_{Hub}=
\begin{cases}
&2 | \bar \eps| (M+1) \qquad (J=0),\\
&2 | \bar \eps| \qquad \;\; \quad (J\neq 0). \\
\end{cases}
\ee
thus illustrating the common origin of the enhancement of the width of the Hubbard bands with the number of orbitals $M$ and of the quasiparticle kinetic energy leading to an enhanced $U_{c}$.\cite{Florens_multiorb}

In ref.~\onlinecite{demedici_Vietri} it was shown numerically within DMFT that a finite Hund's coupling $J$ causes the Hubbard bands to shrink back to values comparable with the single-band case. Here we have analytically shown that the underlying cause coincides with that of the concomitant reduction of quasiparticle kinetic energy, and that this whole phenomenology is due to the number of available hopping channels which in turn is determined by the degeneracy of the ground state and thus ultimately by the degree of symmetry in the model.


\acknowledgements MCh, PVA and LdM are supported by the European Commission through the ERC-StG2016, StrongCoPhy4Energy, GA No724177. MB and MC acknowledge support of italian MIUR through PRIN 2015 (Prot. 2015C5SEJJ001) and PRIN 2017 CenTral. The authors are grateful to Adriano Amaricci, Francesco Grandi and Daniele Guerci for useful discussions.

%

\end{document}